\documentclass[notitlepage,nofootinbib,10pt,pra,twocolumn]{revtex4-1}
\usepackage[colorlinks=true]{hyperref}
\usepackage{amsmath}
\usepackage{amssymb}
\usepackage{color}
\definecolor{deepblue}{rgb}{0,0,0.5}
\definecolor{deepred}{rgb}{0.6,0,0}
\definecolor{deepgreen}{rgb}{0,0.5,0}
\usepackage[mmddyy]{datetime}
\usepackage{wrapfig}
\usepackage{moreverb}
\usepackage{graphicx}
\usepackage{hhline}
\usepackage{listings}
\usepackage{ctable}
\usepackage{moreverb}
\usepackage{algorithm}
\usepackage{algpseudocode}

\newcommand\pythonstyle{\lstset{
language=Python,
basicstyle=\ttm,
numbers=left,
stepnumber=1,
otherkeywords={self},             
keywordstyle=\ttb\color{deepblue},
emph={MyClass,__init__},          
emphstyle=\ttb\color{deepred},    
stringstyle=\color{deepgreen},
frame=tb,                         
showstringspaces=false            %
}}

\lstnewenvironment{python}[1][]
{
\pythonstyle
\lstset{basicstyle=\small,#1}
}
{}

\newcommand\pythoninline[1]{{\pythonstyle\lstinline!#1!}}

\begin{document}

\newcommand{\comb}{\mathrm{comb}}
\newcommand{\rect}{\mathrm{rect}}
\newcommand{\tri}{\mathrm{tri}}
\newcommand{\sinc}{\mathrm{sinc}}
\newcommand{\Gaus}{\mathrm{Gaus}}
\newcommand{\somb}{\mathrm{somb}}
\newcommand{\fstop}{f/\#}
\newcommand{\warn}[1]{{\color{red} \textbf{#1}}}
\newcommand{\blue}[1]{{\color{blue} #1}}
\newcommand{\Ito}{It$\hat{\mathrm{o}}$}

\title{A tensor network discriminator architecture for classification of quantum data on quantum computers}
\author{Michael L.~Wall}
\email{Michael.Wall@jhuapl.edu}
\address{The Johns Hopkins University Applied Physics Laboratory, Laurel, MD 20723, USA}
\author{Paraj Titum}
\address{The Johns Hopkins University Applied Physics Laboratory, Laurel, MD 20723, USA}
\author{Gregory Quiroz}
\address{The Johns Hopkins University Applied Physics Laboratory, Laurel, MD 20723, USA}
\author{Michael Foss-Feig}
\address{Quantinuum, 303 S. Technology Ct., Broomfield, Colorado 80021, USA}
\author{Kaden R.~A.~Hazzard}
\address{Department of Physics and Astronomy, Rice University, Houston, TX 77005-1892, USA}
\address{Rice Center for Quantum Materials, Rice University, Houston, TX 77005-1892, USA}

\begin{abstract}

We demonstrate the use of matrix product state (MPS) models for discriminating quantum data on quantum computers using holographic algorithms, focusing on the problem of classifying a translationally invariant quantum state based on $L$ qubits of quantum data extracted from it.  We detail a process in which data from single-shot experimental measurements are used to optimize an isometric tensor network, the isometric tensors are compiled into unitary quantum operations using greedy compilation heuristics, parameter optimization on the resulting quantum circuit model removes the post-selection requirements of the isometric tensor model, and the resulting quantum model is inferenced on either product state (single-shot measurement) or entangled quantum data.  We demonstrate our training and inference architecture on a synthetic dataset of six-site single-shot measurements from the bulk of a one-dimensional transverse field Ising model (TFIM) deep in its antiferromagnetic and paramagnetic phases.  We find that increasing the bond dimension of the tensor network model, amounting to adding more ancilla qubits to the circuit representation, improves both the average number of correct classifications across the dataset and the single-shot probability of correct classification.  We experimentally evaluate models on Quantinuum's H1-2 trapped ion quantum computer using entangled input data modeled as translationally invariant, bond dimension 4 MPSs across the known quantum phase transition of the TFIM.  Using linear regression on the experimental data near the transition point, we find predictions for the critical transverse field of $h=0.962$ and $0.994$ for tensor network discriminators of bond dimension $\chi=2$ and $\chi=4$, respectively.  These predictions compare favorably with the known transition location of $h=1$ despite training on data far from the transition point.  Our techniques identify families of short-depth variational quantum circuits in a data-driven and hardware-aware fashion and robust classical techniques to precondition the model parameters, and can be adapted beyond machine learning to myriad applications of tensor networks on quantum computers, such as quantum simulation and error correction.

\end{abstract}

\maketitle

\section{Introduction}

Quantum technology continues to advance rapidly.  This progress urgently requires developing and assessing applications of near-term quantum technology and methods to analyze and interpret the large volumes of quantum information these devices will produce.  Beyond-classical computational tasks~\cite{harrow2017quantum} have been demonstrated, including sampling from random unitaries~\cite{arute2019quantum} and boson sampling~\cite{zhong2020quantum}, and smaller-scale demonstrations and proposals exist for quantum simulation~\cite{cirac2012goals}, optimization~\cite{moll2018quantum}, machine learning~\cite{biamonte2017quantum}, and other topics~\cite{montanaro2016quantum}.  In contrast to the widely used randomized benchmarking~\cite{knill2008randomized} and random unitary simulation~\cite{arute2019quantum} use cases, applications like quantum simulation and machine learning will deal with physical data that is expected to be highly structured.  Identifying quantum models that can discover and adapt to this underlying structure, as is the case in successful classical approaches to similar problems~\cite{rawat2017deep,sherstinsky2020fundamentals}, is key to both developing performant algorithms and identifying hardware improvements that improve model performance in relevant metrics.

Tensor networks (TNs), in which a high-rank tensor is decomposed into a contracted network of low-rank tensors, have been used in a variety of applications across physics, chemistry, applied mathematics, and machine learning~\cite{schollwock2011density,cichocki2014tensor,orus2014practical,chan2016matrix,orus2019tensor,dudek2019efficient,torlai2020quantum}.  In quantum physics, TNs are used to represent the coefficients of a wavefunction in a fixed basis, and the topology of the TN is related to the entanglement and correlation structure of the resulting wavefunction~\cite{PhysRevLett.100.070502,eisert2013entanglement}.  Variational methods within the class of TNs define the gold standard for simulations of strongly correlated quantum many-body systems on classical computers, with the most prominent example being the density matrix renormalization group (DMRG) algorithm~\cite{white1992density,schollwock2011density}.  Similar to how classical machine learning has benefitted significantly by tailoring the structure of the learning model to the structure of the data, e.g., to account for its hierarchical or causal nature~\cite{oord2016wavenet}, we can expect that models which encode quantum characteristics of the data may be more performant or interpretable when applied to quantum learning problems.

There is also strong impetus for developing TN models to be employed on quantum computers.  Certain TN architectures naturally define a sequential preparation scheme for quantum states in which qubits holding the quantum state information are correlated using a register of ancilla qubits through a sequence of quantum gates~\cite{huggins2019towards,schon2005sequential}.  The ancilla qubit register is the quantum analog of the contracted indices of the classical TN representation.  However, its effective bond dimension grows exponentially with the number of qubits in the register, leading to a vastly improved scaling of network expressiveness with resources compared to the classical case.  The state of the ancilla qubits in a partially contracted network also carries information about the state, but scales only with the boundary of the contracted region.  This so-called ``holographic" encoding of quantum information~\cite{osborne2010holographic,cirac2011entanglement,kim2017holographic,kim2017noise,kim2017robust} has been utilized in recent works to measure static and dynamic properties of quantum states on quantum computers ~\cite{PhysRevResearch.3.033002,yirka2021qubit,foss2021entanglement,chertkov2021holographic}.  The ability of tensor networks to partially localize information about a quantum state into a reduced-dimensional representation may also bode well for trainability in machine learning applications by avoiding the phenomenon of barren plateaus~\cite{liu2021presence}.

In this work, we explore the use of matrix product states (MPSs), one-dimensional tensor networks, for inferencing from quantum data on quantum computers.  Namely, we focus on the problem of classifying a translationally invariant quantum state given $L$ qubits of data extracted from it.  Our discriminator architecture first generates a prior distribution on the ancilla degrees of freedom in the network, and then conditions the state of the the ancilla distribution on the quantum data to be classified.  A final quantum operation on the ancilla qubits encodes the class label into a subset of those qubits, which can then be read out by measurement in the computational basis.  We detail a process in which the tensors in the network architecture are classically optimized using optimization techniques on manifolds inspired by canonical TN algorithms, the tensors of the optimized classical model are compiled into quantum operations using greedy compilation heuristics~\cite{wall2020Generative,wall2021tree}, and then the resulting model parameters can be fine-tuned based on results obtained from inferencing the quantum model.  In contrast to many other variational quantum circuit approaches, in which a fixed-depth sequence of native gates is optimized over its parameters, our approach only defines the topology of quantum operations between qubits, and allows for an automatic determination of the circuit structure and depth by interfacing with quantum data.

We demonstrate the training of our architecture on a collection of single-shot experimental measurements, in which each shot corresponds to the measurement of $L$ qubits from a fixed set of bases.  Such single-shot measurements can be realized not only in programmable quantum hardware~\cite{PhysRevA.102.012415}, but also in other platforms such as cold atoms~\cite{bakr2009quantum} and trapped ions~\cite{PhysRevLett.100.200502,PhysRevA.103.062614}, and so our training strategy avoids the need for the quantum system providing the data to be programmable.  While the collection of single-shot measurements can be represented as vectors in a $2L$-dimensional space and classically analyzed, the quantum model can capture that these vectors are sparse representations of vectors embedded in a $2^{L}$-dimensional Hilbert space, and use their correlations to define entangled decision boundaries.  In addition, once trained our model can be inferenced on either single-shot measurements or entangled input data, with the latter having no classical analog.  We analyze the performance of both modalities in this work.

This work is organized as follows: Sec.~\ref{sec:MPS} overviews matrix product states, applications to machine learning, and their implementation on quantum hardware; Sec.~\ref{sec:TND} presents our tensor network discriminator architecture and algorithms for classical training, quantum compilation, and other classical preconditioning strategies of the quantum model; Sec.~\ref{sec:Example} describes an application of our model workflow to the one-dimensional (1D) transverse-field Ising model, including classical simulations of model performance and analysis of experiments on Quantinuum's model H1-2 trapped ion quantum computer; finally, Sec.~\ref{sec:Concl} presents conclusions.  Technical details of algorithms used to generate the synthetic training data are presented in the appendices.

\section{Matrix product states}
\label{sec:MPS}

\subsection{Overview}

Matrix product states (MPSs), also known as finitely correlated states, are tensor networks with a one-dimensional topology.  MPSs are of interest because they exactly represent ground states of gapped, local, one-dimensional Hamiltonians~\cite{PhysRevB.73.085115,PhysRevLett.100.070502}, and also because many efficient algorithms have been developed for finding approximate MPS representations of many-body systems, such as the density-matrix renormalization group (DMRG) algorithm~\cite{white1992density}.  The MPS description of a quantum state $|\psi\rangle$ on a  lattice consisting of $L$ $d$-level subsystems is given by the contraction of the $L$ tensors $\{A^{[j]}\}$ as
\begin{align}
\label{eq:MPS}|\psi\rangle&=\sum_{i_0\dots i_{L-1}}\langle l|\mathbb{A}^{[0]i_0}\dots \mathbb{A}^{[L-1]i_{L-1}}|r\rangle |i_0\dots i_{L-1}\rangle\, ,
\end{align}
in which $\mathbb{A}^{[j]i_j}$ denotes the matrix obtained by holding the index $i_j$ fixed in the tensor $A^{[j]}$.  Each tensor ${A}^{[j]}$ is $(\chi, d, \chi)$-dimensional, and the \emph{bond dimension} $\chi$ forms the main convergence parameter of an MPS.  The states $|l\rangle$ and $|r\rangle$ are $\chi$-dimensional states specifying boundary conditions.

The matrix product structure of Eq.~\eqref{eq:MPS} implies that we can place any full-rank $\chi\times\chi$ matrix $\mathbb{X}$ and its inverse $\mathbb{X}^{-1}$ in between any two neighboring tensors in the product without changing the nature of the state.  This leads to a gauge freedom in MPSs that can be exploited to put the tensors into one of a number of so-called canonical forms~\cite{schollwock2011density}.  In this work, a central role will be played by MPSs in the left-canonical form, in which $\sum_{\alpha i}A^{[j]i}_{\alpha\beta}A^{[j]i\star}_{\alpha\beta'}=\delta_{\beta\beta'}$.  Reshaping this tensor into a $\left(\chi d\right)\times \chi$-dimensional matrix, left-canonical form specifies that this is an \emph{isometric matrix} with orthonormal columns.  Hence, this isometric operation can be embedded into a unitary operation $\hat{U}^{[j]}$ such that 
\begin{align}
\label{eq:unitaryembedding} A_{\alpha\beta}^{[j]i_j}=\langle i_j\alpha |\hat{U}^{[j]}| 0\beta\rangle\, ,
\end{align}
and this sequence of unitary operations can then be used to sequentially generate the state as
\begin{align}
\label{eq:psiseq}|\psi\rangle&=  \langle l|_b\hat{U}^{[0]}  |0\rangle_{0} \otimes \dots \otimes  \hat{U}^{[L-1]} |0\rangle_{L-1} \otimes |r\rangle_b\, .
\end{align}
Here, the subscript $b$ denotes states in a $\chi$-dimensional bond Hilbert space, and the other subscripts denote qubit index in a register of $L$ qubits used to hold the generated state.  

Eq.~\eqref{eq:psiseq} codifies a scheme to prepare the state $|\psi\rangle$ using $L+\log_2\chi$ qubits: a register of $L$ qubits plus an additional $\log_2\chi$ qubits to encode the bond degrees of freedom.  However, this scheme can be modified to require only $\log_2\chi+1$ qubits for implementation provided the hardware supports mid-circuit measurement and reuse (MCMR)~\cite{huggins2019towards,pino2020demonstration,gaebler2021suppression}.  In this scheme, $\log_2\chi$ of the qubits form the bond register and the remaining qubit is used to prepare the physical degrees of freedom by sequentially coupling with the bond register.  Then, the physical qubit information is measured before the next physical qubit state is generated, and so this reduced qubit requirement scheme can be used to calculate observables which can be expressed as tensor products of local operators.  This scheme can also be used to generate sample product states in a specified basis, $|i_0\dots i_{L-1}\rangle$ according to their probability of occurrence in the wavefunction, $|\langle \psi|i_0\dots i_{L-1}\rangle|^2$.  The $\log_2\chi$ qubits required are exponentially smaller than the number of classical bits required to implement the MPS, and relatively large system sizes can be amenable to current devices with modest numbers of qubits~\cite{huggins2019towards}.

MPSs can also represent the thermodynamic limit of a translationally invariant system on an infinite chain of sites obtained by the $L\to\infty$ limit of Eq.~\eqref{eq:MPS}, in which case we can capture the state using a single, translationally invariant tensor, dropping the site index superscript in brackets.  This reduced description of an infinite-size state with an MPS tensor is known as an \emph{iMPS}.  Intuitively, far from the boundary the particular boundary vector, e.g., $|r\rangle$ used to generate the state becomes unimportant as the state of the bond degrees of freedom reaches a steady state.  This can be roughly formalized by the fact that the (right) transfer operator of the left-canonical iMPS defined by
\begin{align}
{T}_R\left(A; \mathbb{E}\right)&=\sum_i \mathbb{A}^{i} \mathbb{E} \mathbb{A}^{i\dagger}\, ,
\end{align}
in which $\mathbb{E}$ is a $\chi\times \chi$ matrix, admits the half-infinite reduced density matrix $\hat{\rho}_{\mathrm{half-infinite}}$ obtained by tracing out one half of the infinite chain, as an eigenmatrix with maximal eigenvalue, which is 1 for a normalized iMPS.  Hence, the half-infinite density matrix in bond space is obtained as the steady state of the channel 
\begin{align}
\label{eq:NbBurnIn}\hat{\rho}&=  \mathrm{Tr}_{N_b} \hat{U}  |0\rangle_{N_b} \otimes \dots \mathrm{Tr}_2 \hat{U} |0\rangle_2\otimes  \mathrm{Tr}_1 \hat{U} |0\rangle_1 \otimes |r\rangle_b\, ,
\end{align}
as $N_b\to \infty$, where $\hat{U}$ is the unitary embedding of the MPS tensor $A$ in left-canonical form, defined as in Eq.~\eqref{eq:unitaryembedding}.  In general, tensor networks encode a $D$ dimensional quantum system into a register of qubits on the $(D-1)$-dimensional boundary, so this approach is referred to as ``holographic~\cite{osborne2010holographic,cirac2011entanglement,kim2017holographic,kim2017noise,kim2017robust,PhysRevResearch.3.033002}".   Holographic approaches have been recently exploited in a trapped-ion quantum computer to measure the entanglement entropy defined by the spectrum of $\hat{\rho}_{\mathrm{half-infinite}}$~\cite{yirka2021qubit,foss2021entanglement}, as well as for dynamics~\cite{chertkov2021holographic}.  In practice, since MPSs of finite bond dimension have a finite correlation length, a finite ``burn-in length" $N_b$ suffices (set to $L$ in the results that follow), and the boundary state $|r\rangle$ can be chosen to facilitate rapid convergence to the steady state (see Appendix \ref{app:iMPS} for further algorithmic details).

\subsection{MPSs for machine learning}

Given that MPSs define a sequential preparation procedure for quantum states starting from a given reference state, a natural application of MPSs is generative modeling~\cite{PhysRevX.8.031012,wall2020Generative}.  In this scenario, the tensors $A$ of an MPS are optimized to best represent a collection of training quantum states $\{|\psi_m\rangle\}$, $m=1,\dots,M$ by some metric, such as the average log-likelihood of the data in the MPS
\begin{align}
\mathcal{L}&=\frac{1}{M}\sum_m \ln \left|\langle \psi_m|\psi\left[A\right]\rangle\right|^2\, ,
\end{align}
where $|\psi\left[A\right]\rangle$ denotes the state generated by MPS tensors $A$.  Once trained, the model can be compiled to quantum hardware by finding appropriate embeddings of the tensors $A$ in left canonical form into unitaries as in Eq.~\eqref{eq:unitaryembedding} and then applying the sequential preparation procedure.  A proof of principle of this generative model pipeline has been explored in Ref.~\cite{wall2020Generative}, in which greedy heuristics were applied to compile the isometries into unitaries for hardware with a given gate set and topology using as short of gate sequences as possible.  For a finite-size system, $|r\rangle$ can be chosen to be the vacuum state of the qubits encoding the bond degrees of freedom, while for an infinite-size state the ``burn-in" procedure described above can be enacted to first generate the half-infinite density matrix on the bond states, and then the sequential preparation of a finite portion of the state performed.

\begin{figure}[t]
  \begin{center}
\includegraphics[width=0.95\columnwidth]{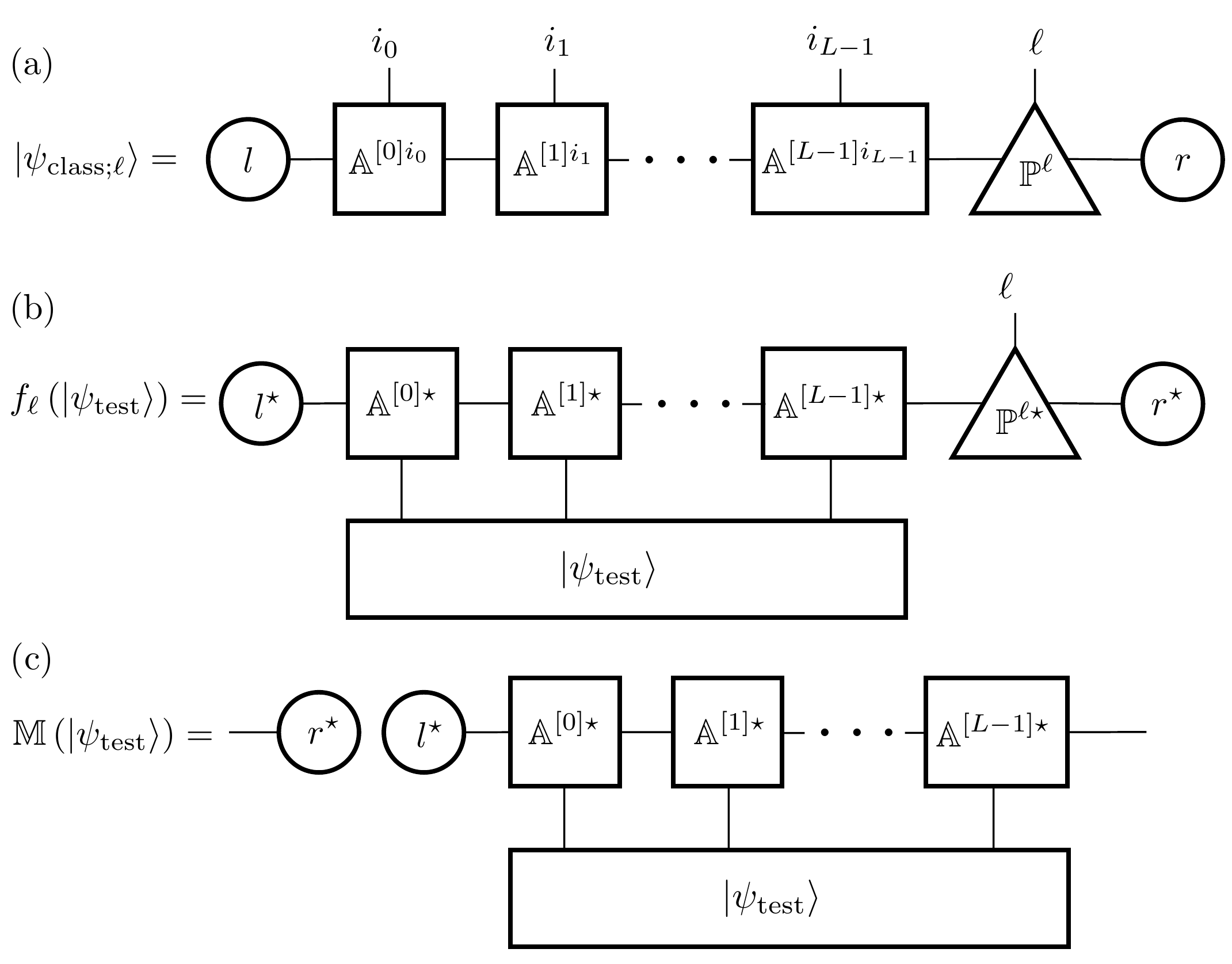}  
\caption{\label{fig:MPSc} \emph{MPS linear classifier model} represented by tensor network graphical notation.  (a) A linear model MPS classifier is obtained by introducing a label tensor $\mathbb{P}^{\ell}$ on one of the bonds of an MPS structure.  (b) The overlap of the MPS classifier wavefunction defined in (a) with a test wavefunction $|\psi_{\mathrm{test}}\rangle$ defines an indicator function $f_{\ell}\left(|\psi_{\mathrm{test}}\rangle\right)$ that can be optimized for classification performance.  (c) The action of the linear MPS model can be interpreted as an encoding of the test state information into the bond space to define an operator $\mathbb{M}\left(|\psi_{\mathrm{test}}\rangle\right)$ shown graphically, followed by classification using the indicator tensor as $\mathrm{argmax}_{\ell} \mathrm{Tr}\left(\mathbb{P}^{\ell\dagger}\mathbb{M}\left(|\psi_{\mathrm{test}}\rangle\right) \right)$.}
\end{center}
\end{figure}

If we instead consider a supervised discriminative task, we now wish to classify a quantum state $|\psi_{\mathrm{test}}\rangle$ into one of $N_C$ classes given a collection of training states and class labels $\mathcal{T}=\{(|\mathbf{x}_m\rangle, \ell_m),m=1,\dots M\}$.  One means of doing so is to define an MPS linear model~\cite{stoudenmire2016supervised} by introducing a $\chi\times N_C\times \chi$-dimensional label tensor $\mathbb{P}^{\ell}$ in between the bonds of the MPS tensors in the representation Eq.~\eqref{eq:MPS} to define a classifier state $|\psi_{\mathrm{class};\ell}\rangle$ as
\begin{align}
|\psi_{\mathrm{class};\ell}\rangle&=\sum_{i_0\dots i_{L-1}}\langle l|\mathbb{A}^{[0]i_0}\dots \mathbb{A}^{[L-1]i_{L-1}}\mathbb{P}^{\ell}|r\rangle| i_0\dots i_{L-1}\rangle\, .
\end{align}
This object is shown graphically in Fig.~\ref{fig:MPSc}(a).  The overlap of this classifier MPS with a test quantum state $|\psi_{\mathrm{test}}\rangle$ defines an indicator function $f_{\ell}\left(|\psi_{\mathrm{test}}\rangle\right)$ as
\begin{align}
\nonumber &f_{\ell}\left(|\psi_{\mathrm{test}}\rangle\right)=\langle \psi_{\mathrm{class};\ell}| \psi_{\mathrm{test}}\rangle =\\
&\sum_{i_0\dots i_{L-1}}\Big(\langle l|\mathbb{A}^{[0]i_0}\dots \mathbb{A}^{[L-1]i_{L-1}}\mathbb{P}^{\ell}|r\rangle\Big)^{\star}\langle i_0\dots i_{L-1}|\psi_{\mathrm{test}}\rangle \, ,
\end{align}
shown graphically in Fig.~\ref{fig:MPSc}(b).  This indicator function can then be optimized over the tensors of the MPS ($A$ and $P$) such that $f_{\ell}\left(|\psi_{\mathrm{test}}\rangle\right)\approx 1$ when $|\psi_{\mathrm{test}}\rangle$ is in class $\ell$ and $f_{\ell'}\left(|\psi_{\mathrm{test}}\rangle\right)\approx0$ for all other classes $\ell'$, with performance averaged over the training set $\mathcal{T}$.  We note that the position of the indicator tensor $\mathbb{P}^{\ell}$ in the MPS chain can be shifted using gauge freedom, and so we have shifted it to the right end for convenience.  We can re-write the indicator function as $f_{\ell}\left(|\psi_{\mathrm{test}}\rangle\right)=\mathrm{Tr}\left(\mathbb{P}^{\ell\dagger}\mathbb{M}\left(|\psi_{\mathrm{test}}\rangle\right) \right)$ in which \
\begin{align}
\nonumber \mathbb{M}\left(|\psi_{\mathrm{test}}\rangle\right)&= \sum_{i_0\dots i_{L-1}}\mathbb{A}^{[L-1]i_{L-1}\dagger}\dots \\
\label{eq:MPSClassMdef}&\times \mathbb{A}^{[0]i_0\dagger }\langle i_0\dots i_{L-1}|\psi_{\mathrm{test}}\rangle|l\rangle\langle r|\, .
\end{align}
$\mathbb{M}\left(|\psi\rangle\right)$ is shown graphically in Fig.~\ref{fig:MPSc}(c), and represents the overlap between the vector $|\psi\rangle$ and the state represented by the MPS tensors $A$ in bond space.  Hence, the indicator tensor $\mathbb{P}^{\ell}$ performs classification based on the bond space encoding of this overlap.  In this way, an MPS classifier performs a lower-dimensional ``holographic" encoding of the information needed for classification into the bond degrees of freedom, in analogy to how classical encoders produce reduced dimensional representations in an appropriately defined latent space.

This observation leads to a duality between generative and discriminative MPS models.  Namely, consider that we have two generative MPS models with tensors $A$ and $A'$ and bond dimensions $\chi$ and $\chi'$ encoding training sets $\mathcal{T}$ and $\mathcal{T}'$ with high fidelity.  If we now consider that the elements of $\mathcal{T}$ are in class 0 and those of $\mathcal{T}'$ are in class 1, we can construct an MPS classifier with bond dimension $\chi_c=\chi+\chi'$ as
\begin{align}
\nonumber |\psi_{\mathrm{class};\ell}\rangle&=\mathcal{N}\sum_{i_0\dots i_{L-1}}  |i_0\dots i_{L-1}\rangle\langle l|\otimes \langle l'| \\
\nonumber &\times \left(\mathbb{A}^{[0]i_0}\otimes \mathbb{I}_{\chi'}+\mathbb{I}_{\chi}\otimes \mathbb{A}^{'[0]i_0}\right)\\
\nonumber & \times \dots \left(\mathbb{A}^{[L-1]i_{L-1}}\otimes \mathbb{I}_{\chi'}+\mathbb{I}_{\chi}\otimes \mathbb{A}^{'[L-1]i_{L-1}}\right)  \\
\label{eq:MPSc}&\times\left(\delta_{\ell,0}\hat{\mathcal{P}}_{0}+\delta_{\ell,1}\hat{\mathcal{P}}_{1}\right)|r\rangle\otimes |r'\rangle\, ,
\end{align}
in which $\delta_{\ell,\ell'}$ is the Kronecker delta, $\hat{\mathcal{P}}_0$ ($\hat{\mathcal{P}}_1$) is a projector onto the bond states connecting to the tensors $A$ ($A'$), and $\mathcal{N}$ is a normalization factor.  The indicator function associated with this MPS model applied to a test state $|\mathbf{x}\rangle$ is
\begin{align}
f_{\ell}\left(|\mathbf{x}\rangle\right)&=\mathcal{N}\left(\delta_{\ell,0}\langle \psi\left[A\right]|\mathbf{x}\rangle+\delta_{\ell,1}\langle \psi\left[A'\right]|\mathbf{x}\rangle\right)\, .
\end{align}
Hence, this indicator function behaves as desired, selecting the class according to the state with higher overlap, and will be accurate so long as the two classes are sufficiently distinguishable (have small overlap $\ll 1$).  In the extreme case that one of the generative models is provided as a test state, the correct class will have weight $\propto 1$ while the incorrect class will be weighted by a factor $\propto \langle \psi\left[A\right]|\psi\left[A'\right]\rangle$ that will be $\ll1$ when the classes are distinguishable.

The above process generates an accurate classifier provided that the training data within each class can be well-described as an MPS with the specified bond dimension.  However, there are several drawbacks of this approach.  Perhaps most strikingly, this process requires building of generative models which capture all properties of the training data, and then requires all of this information from all classes to be encoded into the classifier model.  This leads to potentially very large bond dimensions.  However, just as in classical learning, it is likely that the classification decision can be made based on a set of features derived from the data, whose size is much smaller than the size of the full dataset.  In the architecture we present in the next section, we avoid this drawback by training a generator in feature space and the discriminator simultaneously.  This enables us to model only the parts of the data necessary for the classification decision, reducing the resource requirements for the model.  Further, this approach does not require that the training data has an efficient MPS representation, but only that the set of features used for the classification decision does.

The second point we will address with our tensor network discriminator architecture has to do with the implementation of a classifier like Eq.~\eqref{eq:MPSc} on quantum hardware.  As follows from Eqs.~\eqref{eq:MPSc} and \eqref{eq:psiseq}, an MPS classifier can be applied to a separable feature state $|\mathbf{x}\rangle=|x_0\rangle\otimes\dots\otimes |x_{L-1}\rangle$ using $\log_2\chi+\lceil\log_2 N_C\rceil$ qubits and MCMR as
\begin{align}
\label{eq:discrimcircuit}\langle r|_{b}\hat{\mathcal{U}}|0\rangle_c  \langle 0|\hat{U}^{[L-1]\dagger} |x_{L-1}\rangle \otimes \dots \otimes \langle 0|\hat{U}^{[0]\dagger}  |x_0\rangle \otimes |l\rangle_b\, ,
\end{align}
in which $\hat{U}^{[j]}$ is a unitary embedding of the MPS tensor defining the generative interpretation of the model Eq.~\eqref{eq:unitaryembedding} and $\hat{\mathcal{U}}$ is a similar unitary embedding of $\mathbb{P}^{\ell}$ such that
\begin{align}
\label{eq:mcUdef}P_{\alpha\beta}^{\ell}&=\langle \ell\alpha |\mathcal{U}| 0\beta\rangle\, ,
\end{align}
in which $\ell$ is the class label interpreted in binary.  Following several executions of the circuit, the most probable inferred value of $\ell$ obtained by measurements in the computational basis can be assigned as the predicted class label.  

As is evident from Eq.~\eqref{eq:discrimcircuit}, this circuit also requires postselection on the physical qubits being in the state $|0\rangle$ following action of $\hat{U}^{[j]}$.  When the state $|1\rangle$ is instead measured it indicates that the action of the circuit does not correspond to action of the tensor $A^{[j]}$ defining the MPS, but rather to the portion of the unitary embedding $\hat{U}^{[j]}$ that is not determined uniquely by the MPS representation.  While the probability of successful postselection may be high in the case that the training and test feature vectors have significant overlap and the MPS model is very expressive, it is clearly preferred to develop a model that does not require these conditions to be met.  In the next section we develop an architecture inspired by these observations that will obviate the need for postselection.

\section{Tensor network discriminator architecture}
\label{sec:TND}

\begin{figure*}[t]
  \begin{center}
\includegraphics[width=1.8\columnwidth]{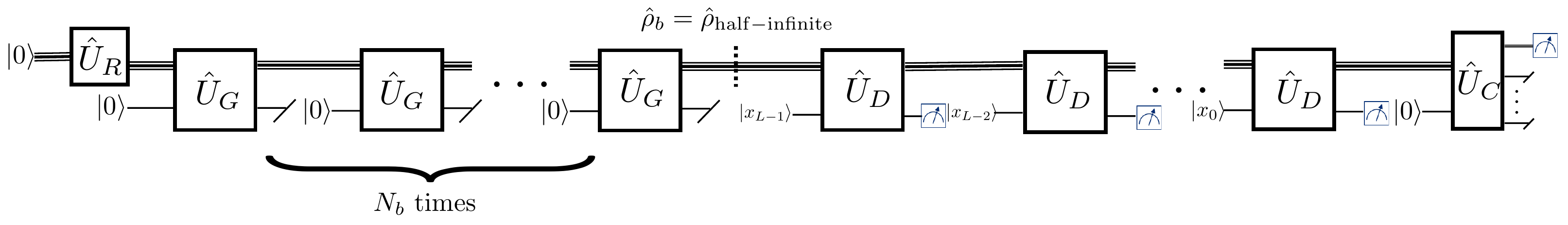}  
\caption{\label{fig:PSFig} \emph{Diagram of tensor network discriminator architecture}, with operation read left to right as in quantum circuit diagrams.  First, the operator $\hat{U}_R$ creates an initial state in the $\log_2\chi$ bond qubits, starting from the vacuum state.  Next, this bond state is coupled with a physical qubit via the operator $\hat{U}_G$ and the physical qubit state discarded and reset $N_b$ times.  This creates a bond state that can be interpreted as a feature space representation of the density matrix obtained by tracing out half of the infinite system.  This bond state is now combined with the input state $|\mathbf{x}\rangle$ a single qubit at a time via the operator $\hat{U}_D$, and the physical qubit measured and re-initialized after each application.  For clarity, this operation is drawn as though the test state is a product state, but the process also applies to any entangled input state stored in an $L$-qubit register.  Finally, the resulting bond degrees of freedom which have been conditioned on the observation of the state $|\mathbf{x}\rangle$ are operated on by $\hat{U}_C$, and the measurement of $\log_2N_C$ of these qubits defines a binary representation of the predicted class label.}
\end{center}
\end{figure*}

In this section, we put forth a tensor network discriminator architecture for quantum hardware that enables efficient classical training and preconditioning, avoids the need for postselection that arises from naive application of an MPS model for classification, and produces local cost functions with a data-defined circuit architecture amenable to refinement and optimization with the quantum device in the loop.  While our methodology can also be straightforwardly adapted to finite-size systems with boundaries, we will henceforth specialize to the case of learning from a finite string sampled from an infinite, translationally invariant state.

Our discriminator consists of four key model components to be trained, which we will denote as $\hat{U}_R$, $\hat{U}_G$, $\hat{U}_D$, and $\hat{U}_C$.  $\hat{U}_R$ is a $\chi\times \chi$ unitary operator acting on the bond degrees of freedom to form an appropriate representation of the right boundary condition $|r\rangle$.  $\hat{U}_G$ is a $\left(\chi d\right)\times \left(\chi d\right)$-dimensional unitary that generates an appropriate prior distribution for the feature space in the bond degrees of freedom starting from the state generated by $\hat{U}_R$.  Together, $\hat{U}_R$ and $\hat{U}_G$ generate the appropriate boundary conditions and prior information for the classifier, similar to the contribution $|l\rangle\langle r|$ in the matrix $\mathbb{M}$ defined in Eq.~\eqref{eq:MPSClassMdef}.  $\hat{U}_D$ is a $\left(\chi d\right)\times \left(\chi d\right)$-dimensional unitary that conditions the state of the bond degrees of freedom based on an input vector $|\mathbf{x}\rangle$.  Finally, $\hat{U}_C$ is a $\left(\chi d\right)\times\left(\chi d\right)$-dimensional\footnote{$\hat{U}_C$ is not required to act on the physical qubit; however, since the physical qubit is needed for other parts of the architecture we have included it as a potential resource to be utilized by $\hat{U}_C$.} unitary that encodes predicted class labels into the probabilities of the states in a $\log_2N_C$-dimensional register based on the state of the bond degrees of freedom output from $\hat{U}_D$.  In this way, $\hat{U}_G$ is analogous to the unitaries $\hat{U}$ generating an MPS state as in Eqs.~\eqref{eq:unitaryembedding} and \eqref{eq:psiseq}, $\hat{U}_D$ is analogous to the adjoints of $\hat{U}$ utilized in an MPS classifier as Eq.~\eqref{eq:discrimcircuit}, and $\hat{U}_C$ classifies based on the state of the bond degrees of freedom analogous to $\hat{\mathcal{U}}$ in Eqs.~\eqref{eq:discrimcircuit} and \eqref{eq:mcUdef}.  In addition to the input parameters $L$ (the number of qubits encoding the input data) and $N_C$ (the number of classes) there are two hyperparameters of the model: $N_b$ is the number of times $\hat{U}_G$ is applied to reach a steady-state bond register prior distribution and $\chi$ is the bond dimension.

This process is displayed visually in Fig.~\ref{fig:PSFig} for the case of a product input vector $|\mathbf{x}\rangle$.  The total qubit requirements for the classifier, assuming MCMR, are $\log_2\chi+1$.  This does not include the resources required to encode the input vector $|\mathbf{x}\rangle$.  In the general case that this is an arbitrary quantum state it will require $L$ qubits to encode.  Two important special cases are (1) when all input vectors are product states, in which case the physical qubit used in the classifier can be utilized and no additional resources are required, and (2) when the input data is described by an MPS with bond dimension $\chi'$, in which case an additional $\log_2\chi'$ qubits are required.  We will demonstrate both of these modalities in Sec.~\ref{sec:Example}.

\subsection{Classical training and preconditioning of tensor network architecture}
\label{sec:classtraining}
We now turn to how the tensor network discriminator architecture can be optimized classically using adaptations of standard tensor network techniques.  This classical preconditioning serves several benefits.  For one, the classical preconditioning will define a quantum circuit architecture with initial guesses that can then be optimized further on the quantum device, with the initialization helping to avoid local minima and barren plateaus~\cite{mcclean2018barren,grant2019initialization,cerezo2021cost}.  In stark contrast to most other approaches for variational circuit learning, in which a predefined circuit architecture is optimized or operations are characterized as general dense unitaries without reference to their representation in native gates for a given hardware realization, our approach will produce a family of quantum circuits whose depth and native gate content are discovered from the correlation structure of the training data itself.  Finally, from a benchmarking perspective this process produces models that can be analyzed on either classical or quantum hardware to assess the quality of noise models for the latter in structured use cases. 

The operators $\hat{U}_R$, $\hat{U}_G$, $\hat{U}_D$, and $\hat{U}_C$ can be interpreted as unitary embeddings of tensors $R$, $G$, $D$, and $C$, analogous to the embeddings described in Sec.~\ref{sec:MPS}.  $R$ describes a normalized $\chi$-dimensional state, and so is an element of the sphere manifold of dimension $\chi$, $S^{\chi}$, whose elements have unit norm\footnote{Throughout, the term norm refers to the $L_2$-norm.}, and the unitary embedding takes the form $R_{\alpha}=\langle \alpha|\hat{U}_R|0\rangle$.  The tensor $G$ can be described as a $\left(\chi,d,\chi\right)$-dimensional tensor in left canonical form, such that the $\left(\chi d\right)\times \chi$-dimensional matrix reshaping of this tensor is an element of the Stiefel manifold $\mathrm{St}\left(\chi d,\chi\right)$ of matrices with orthonormal columns.  Similarly, $D$ is a $\left(\chi,d,\chi\right)$-dimensional tensor where the adjoint of the $\chi\times \left(\chi d\right)$-dimensional matrix reshaping is an element of $\mathrm{St}\left(\chi d,\chi\right)$.  The unitary embeddings of these tensors are analogous to Eq.~\eqref{eq:unitaryembedding} for MPS tensors.  Finally, $C$ is a $\chi\times N_C$ dimensional matrix that is an element of $\mathrm{St}\left(\chi,N_C\right)$ and embedded as $C_{\alpha,\ell}=\langle 0|_b\langle \ell|\hat{U}_C|\alpha\rangle_b|0\rangle$, where we have assumed binary classification for simplicity.

Given that all of the tensors in our architecture are elements of Riemannian manifolds, using manifold optimization techniques is a natural approach.  Several recent works have discussed manifold optimization techniques, including within the context of tensor networks and other quantum applications~\cite{luchnikov2020riemannian,hauru2021riemannian}.  In order to apply these techniques, we need to define a cost function for the optimization over a training set $\mathcal{T}=\{|\mathbf{x}_m\rangle, \ell_m,m=1,\dots M\}$.  The key output to be optimized is the density matrix $\rho$ of the class qubit register following the process outlined in Fig.~\ref{fig:PSFig}.  Namely, we would like to optimize $\rho$ such that the elements $\rho_{\ell_m\ell_m}$, corresponding to the probability of measuring $|\ell_m\rangle$ when inferencing a state $|\mathbf{x}_m\rangle$, are as large as possible, while all other probabilities $\rho_{\bar{\ell}\bar{\ell}}$ are as small as possible, averaged over the training set. When evaluated with the tensors $R$, $G$, etc. rather than their unitary embeddings, the analog of this density matrix will no longer have unit trace, and so is not a true density matrix.  Hence, in order to treat all training states on the same footing, this object should be divided by its trace when evaluated in this fashion.  In practice, we find that minimizing the cost function $\frac{1}{M}\sum_m \mathcal{C} \left(\rho^m,\ell_m\right)$, in which 
\begin{align}
\mathcal{C}\left(\rho,\ell_m\right)&=\frac{\sum_{\ell} \rho_{\ell\ell}-2\rho_{\ell_m\ell_m}}{\sum_{\ell} \rho_{\ell\ell}}\, ,
\end{align}
performs well.  We note that the coherences (off-diagonal elements) of the density matrix play no role in the optimization, since the final readout is in the computational basis.  

\begin{figure*}[t]
  \begin{center}
\includegraphics[width=1.8\columnwidth]{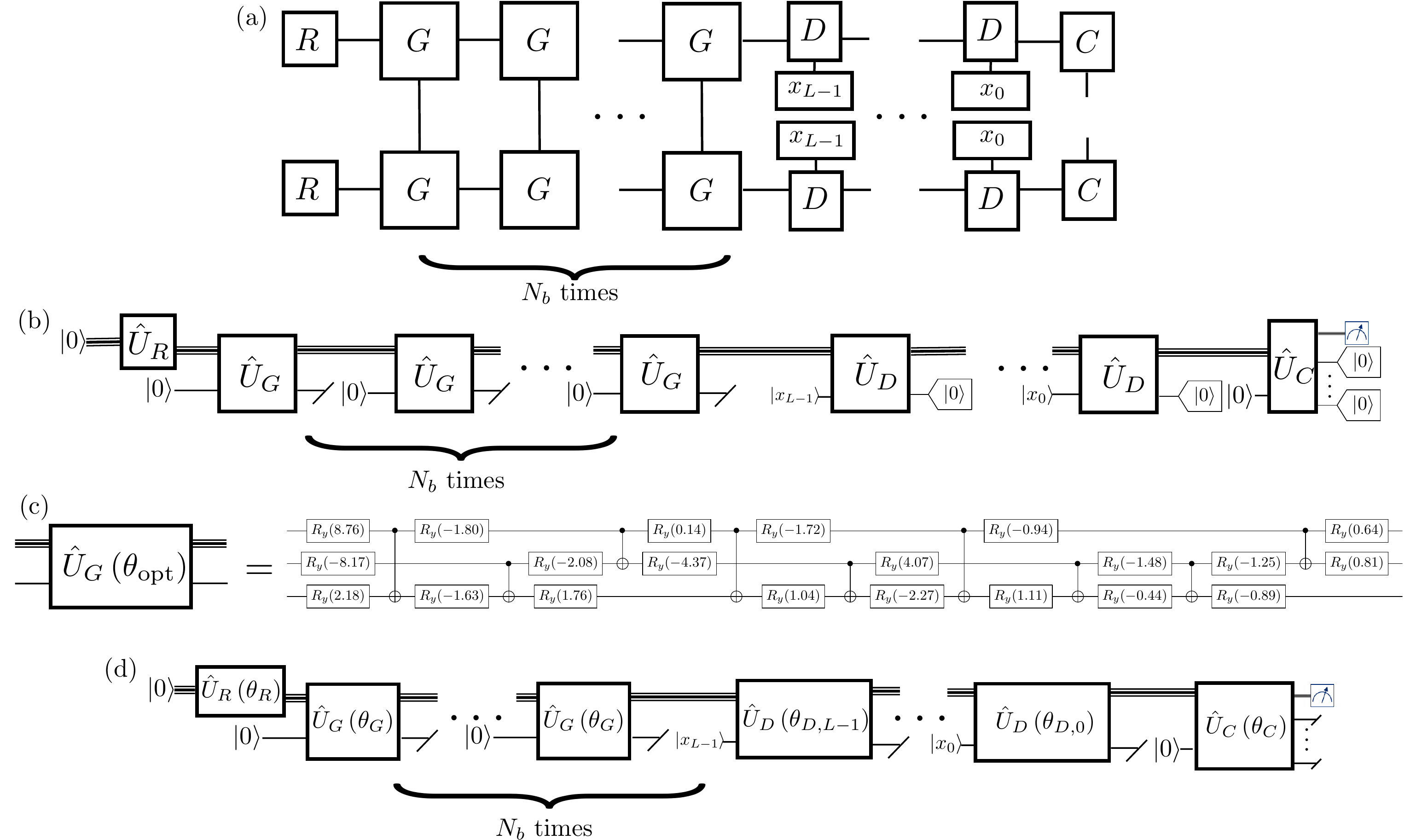}  
\caption{\label{fig:CTFig} \emph{Training of discriminator architecture}.  (a) Tensor network diagram for the construction of $\rho_{\ell\ell}$ from the tensors $R$, $G$, $D$, and $C$, as in algorithm \ref{alg:CFtensor}.  (b) Compilation of the unitary embeddings of $R$, $G$, $D$, and $C$ results in a quantum circuit sampling from the distribution $\rho_{\ell,\ell}$ with postselection in the $|0\rangle$ state indicated by tags.  (c) Example compiled unitary for $\hat{U}_G$ evaluated at the optimal parameters for the postselected circuit in (b), here shown with 2 bond qubits.  Gates connecting qubit wires are CNOTs, and the single-qubit rotations have the matrix representation $\hat{R}_y\left(\theta\right)\equiv \left(\begin{array}{cc} \cos\frac{\theta}{2} &-\sin\frac{\theta}{2} \\ \sin\frac{\theta}{2}&\cos\frac{\theta}{2}\end{array}\right)$.  (d) Circuit ansatz for model without postselection, in which the compiled circuit architectures are the same as the postselected model (e.g., panel (c) for $\hat{U}_G$), but with different values for the parameters.}
\end{center}
\end{figure*}

Summarizing, we can define a cost function amenable to manifold optimization over the tensors $R$, $G$, $D$, $C$ as in Algorithm \ref{alg:CFtensor}, which assumes a training set of product state vectors $|\mathbf{x}_m\rangle=|x_{m,0}\rangle\dots |x_{m,L-1}\rangle=\left(\sum_{i_0}x_{m,0}^{i_0}|i_0\rangle\right)\dots \left(\sum_{i_{L-1}}x_{m,{L-1}}^{i_{L-1}}|i_{L-1}\rangle\right)$ for simplicity.  As discussed further in Sec.~\ref{sec:Example}, product state samples correspond to the case of single-shot measurements relevant for near-term experiments.
\begin{algorithm}[H]
\caption{Cost function for classification accuracy of product-state training data over tensor elements \label{alg:CFtensor}}
\begin{algorithmic}
\Require Input tensors $G_m,D_m\in \mathrm{St}\left(\chi d,\chi\right)$, $C\in\mathrm{St}\left(\chi,N_C\right)$, $R\in S^{\chi}$.
\Require Training set of product states $\mathcal{T}=\{|\mathbf{x}_m\rangle, \ell_m,m=1,\dots M\}$
\Require Hyperparameters $N_b\ge 1$, $\chi\ge1$.
\Function{ClassificationCost}{$R,G,D,C,N_b,\chi,\mathcal{T}$}
\State $G\gets \mathrm{Reshape}(G_m,(\chi,d,\chi))$
\State $D\gets \mathrm{Reshape}(D_m^{\dagger},(\chi,d,\chi))$
\State $B_{\alpha}^i\gets \sum_{\beta}G_{\alpha \beta}^{i}R_{\beta}$
\State $V_{\alpha,\alpha'}\gets \sum_i B_{\alpha}^{i\star}B_{\alpha'}^{i}$
\State $N\gets 1$
\While{$N < N_b'$}
\State $B_{\alpha\beta}^{i} \gets \sum_{\alpha'} V_{\alpha\alpha'}G_{\alpha' \beta}^i$
\State $V_{\alpha \alpha '}\gets \sum_{\beta,i} B_{\beta \alpha'}G_{\beta \alpha}^{i\star}$
\State $N\gets N+1$
\EndWhile
\State $B_{\alpha\beta}^{m} \gets \sum_{i} x_{m,L-1}^{i}D_{\alpha\beta}^i$
\State $Q_{\alpha\beta}^m\gets \sum_{\gamma}B_{\alpha\gamma}^m V_{\gamma\beta}$
\State $V_{\alpha\beta}^m\gets \sum_{\gamma}Q_{\alpha,\gamma}^mB_{\beta\gamma}^{m\star}$
\State $N\gets 2$
\While{$N < L$}
\State $B_{\alpha\beta}^{m} \gets \sum_{i} x_{m,L-N}^{i}D_{\alpha \beta}^i$
\State $Q_{\alpha\beta}^m\gets \sum_{\gamma}B_{\alpha\gamma}^m V_{\gamma\beta}^m$
\State $V_{\alpha\beta}^m\gets \sum_{\gamma} Q_{\alpha,\gamma}^mB_{\beta\gamma}^{m\star}$
\State $N\gets N+1$
\EndWhile
\State $\rho_{\ell\ell}^m\gets \sum_{\alpha\beta} C_{\alpha\ell} V_{\alpha\beta}^mC_{\beta\ell}^{\star}$
\State \Return $\frac{1}{M} \sum_m \mathcal{C}\left(\rho^m,\ell_m\right)$
\EndFunction
\end{algorithmic}
\end{algorithm}
A tensor network diagram for the construction of $\rho_{\ell,\ell'}$ is shown in Fig.~\ref{fig:CTFig}(a).  In practice, we achieve the best performance when we do a first round of optimization jointly across all tensors followed by rounds of optimization where a single tensor is optimized with all others held fixed.  We also note that high-quality software packages for Riemannian manifold optimization are available in a variety of programming languages~\cite{boumal2014manopt,townsend2016pymanopt}.

With the optimized tensors in hand, we move to compiling the embedded unitaries to quantum circuits for deployment to quantum hardware.  As in past work~\cite{wall2020Generative,wall2021tree}, we utilize a greedy iterative heuristic for compilation in which progressively deeper circuits are constructed from past iterations until the distance between the compiled unitary and the unitary embedding of the tensor over the elements where the tensor is defined is less than a defined tolerance.  The method is initialized with an ansatz consisting of single-qubit rotations on each qubit, and updates to previous iterations consist of a single, elemental two-qubit entangling gate (e.g., CNOT) together with single-qubit rotations on the modified qubits.  At each iteration the circuits are optimized over their parameters (e.g., the rotation angles of the single-qubit rotations) and the best performing $h$ circuits are kept to initialize the next iteration, with $h$ being a hyperparameter.  This procedure can be readily adapted to restrictions in hardware connectivity, can be modified to account for hardware noise through the cost function used for model downselection, and utilizes the freedom in compiling the isometric tensors to unitaries to discover compiled gate sequences that are as short as possible.  

As also described in Ref.~\cite{wall2020Generative}, even when the MPS tensors are in a properly defined canonical form there is still a unitary gauge freedom on the bond degrees of freedom that can be exploited to reduce the complexity of the compiled unitary embedding.  In the finite-size systems studied in Ref.~\cite{wall2020Generative}, a ``diagonal gauge" was put forwards in which a permutation operator was applied on the bond degrees of freedom to attempt to maximize diagonal dominance of the compiled unitary in the computational basis.  In this work we define an analog of this diagonal gauge for iMPSs.  Namely, we optimize a general unitary rotation on the bond degrees of freedom $\mathbb{W}$ over the cost function
\begin{align}
\sum_{\alpha,\beta} \Delta_{\alpha\beta} \left[\sum_i \tilde{G}_{\alpha \beta}^{i}\tilde{G}_{\alpha \beta}^{i\star}+ \tilde{D}_{\alpha \beta}^{i}\tilde{D}_{\alpha \beta}^{i\star}\right]\, ,
\end{align}
in which $\Delta_{\alpha,\beta}=\sum_{\mu} \left|b_{\mu}\left(\alpha\right)-b_{\mu}\left(\beta\right)\right|^2$ is a matrix of differences between the binary vector representations $\mathbf{b}\left(\gamma\right)$ (i.e., Hamming distance) of the bond indices that penalizes off-diagonal terms and
\begin{align}
\tilde{G}_{\alpha\beta}^{i}=\sum_{\alpha'\beta'}W_{\alpha\alpha'}G_{\alpha'\beta'}^{i} W_{\beta\beta'}^{\star}\, ,\\
\tilde{D}_{\alpha\beta}^{i}=\sum_{\alpha'\beta'}W_{\alpha\alpha'}D_{\alpha'\beta'}^{i} W_{\beta\beta'}^{\star}\, .
\end{align}
As above, we optimize this cost function using Riemannian optimization over the elements $\mathbb{W}\in\mathrm{St}\left(\chi,\chi\right)$.  Following this optimization, the tensors $R$, $G$, $D$, and $C$ have their bond degrees of freedom transformed using this unitary operator.  An example gauged and compiled circuit using our compilation procedure is shown in Fig.~\ref{fig:CTFig}(c) using CNOTs and $R_y$ rotations.

The compiled unitaries embedding the tensors $R$, $G$, $D$, and $C$ immediately define a quantum circuit which produces samples of the binary encoding of the class label $\ell$ according to the distribution $\rho_{\ell,\ell}$.  However, because of the isometric nature of the tensors $R$, etc., this circuit requires postselection of the physical qubit from $\hat{U}_D$ and the bond qubits from $\hat{U}_C$, as shown in Fig.~\ref{fig:CTFig}(b).  We can remedy this by replacing all instances of postselection by a trace over the corresponding qubits; however, the parameters of the model have been optimized for the postselected operation by virtue of the isometric nature of the tensors $R$, etc., and using these same parameters directly without postselection usually results in poor performance.  One way to improve performance is to treat each of the operators, e.g., $\hat{U}_G$, as a general unitary and optimize its parameters for the circuit without postselection.  Such an approach has been taken in Ref.~\cite{huggins2019towards}, where the unitaries were parameterized as the exponential of a Hermitian matrix and the elements of this Hermitian matrix optimized by the simultaneous perturbation stochastic approximation (SPSA).  However, such an approach will produce dense unitaries which then have to be compiled to quantum hardware, generally requiring deep circuits~\cite{preskill2001lecture}.  In contrast, our approach produces underconstrained isometries, and the freedom in embedding these isometries into unitaries can be used to produce shorter circuits accounting for hardware constraints.

Our method to improve the performance of the model architecture when applied without postselection is to keep the circuit architecture that has been learned for operation with postselection [e.g., the circuit compiling the unitary embedding $\hat{U}_G$ in Fig.~\ref{fig:CTFig}(c)], and to directly optimize the parameters of this circuit for a cost function using operation without postselection.  This leads to the circuit architecture in Fig.~\ref{fig:CTFig}(d) the with angles $\theta$ as the parameters to be optimized.  As the operation of this circuit is now a completely-positive trace preserving (CPTP) map, $\sum_{\ell} \rho_{\ell\ell}=1$.  In order to balance the optimization across many training instances in the dataset, we find it is useful to use the regularized cost function of Ref.~\cite{huggins2019towards}
\begin{align}
\mathcal{C}_{\lambda,\eta}\left(\rho,\ell_m\right)&=\max\left(\rho_{\bar{\ell}_m\bar{\ell}_m}-\rho_{\ell_m\ell_m}+\lambda,0\right)^{\eta}\, ,
\end{align}
in which $\lambda$ and $\eta$ are hyperparameters and $\bar{\ell}_m$ is the label of the incorrect class for training vector $|\mathbf{x}_m\rangle$ with highest model probability.  This results in algorithm \ref{alg:CFNoPS} for the classification accuracy cost given a set of circuit parameters $\boldsymbol{\theta}=\left\{\theta_R,\theta_G,\{\theta_{D,i},i=0,\dots,L-1\}, \theta_C\right\}$.
\begin{algorithm}[H]
\caption{Cost function for classification accuracy of product-state training data without postselection \label{alg:CFNoPS}}
\begin{algorithmic}
\Require Input parameterized gate sequences $\hat{U}_R\left(\bullet\right)$, $\hat{U}_G\left(\bullet\right)$, $\hat{U}_D\left(\bullet\right)$, $\hat{U}_C\left(\bullet\right)$ and initial guesses for the parameters $\boldsymbol{\theta}=\left\{\theta_R,\theta_G,\{\theta_{D,i},i=0,\dots,L-1\}, \theta_C\right\}$.
\Require Training set of product states $\mathcal{T}=\{|\mathbf{x}_m\rangle, \ell_m,m=1,\dots M\}$
\Function{ClassificationCostNoPostselection}{$U_R,U_G,U_D,U_C,N_b,\boldsymbol{\theta},\mathcal{T}$}
\State $B_{\alpha}^i\gets \sum_{\beta}U_{G;\left(\alpha,i\right)\left(\beta,0\right)}\left(\theta_G\right)U_{R;\beta,0}\left(\theta_R\right)$
\State $V_{\alpha,\alpha'}\gets \sum_i B_{\alpha}^{i\star}B_{\alpha'}^{i}$
\State $N\gets 1$
\While{$N < N_b'$}
\State $B_{\alpha\beta}^{i} \gets \sum_{\alpha'} V_{\alpha\alpha'}U_{G;\left(\alpha',i\right)\left(\beta,0\right)}\left(\theta_G\right)$
\State $V_{\alpha \alpha'}\gets \sum_{\beta,i} B_{\beta \alpha'}U_{G;\left(\alpha,i\right),\left(\beta,0\right)}^{\star}\left(\theta_G\right)$
\State $N\gets N+1$
\EndWhile
\State $B_{\alpha\beta}^{mp} \gets \sum_{i} x_{m,L-1}^{i}U_{D;\left(\alpha,p\right)\left(\beta,i\right)}\left(\theta_{D,L-1}\right)$
\State $Q_{\alpha\beta}^{mp}\gets \sum_{\gamma}B_{\alpha\gamma}^{mp} V_{\gamma\beta}$
\State $V_{\alpha\beta}^m\gets \sum_{\gamma p} Q_{\alpha,\gamma}^{mp}B_{\beta\gamma}^{mp\star}$
\State $N\gets 2$
\While{$N < L$}
\State $B_{\alpha\beta}^{mp} \gets \sum_{i} x_{m,L-N}^{i}U_{D;\left(\alpha,p\right)\left(\beta,i\right)}\left(\theta_{D,L-N}\right)$
\State $Q_{\alpha\beta}^{mp}\gets \sum_{\gamma}B_{\alpha\gamma}^{mp} V_{\gamma\beta}^m$
\State $V_{\alpha\beta}^{m}\gets \sum_{\gamma,p}Q_{\alpha,\gamma}^{mp}B_{\beta\gamma}^{mp\star}$
\State $N\gets N+1$
\EndWhile
\State $\rho_{\ell\ell}^m\gets \sum_{\alpha\beta,\gamma} U_{C;\left(\gamma,\ell\right),\left(\alpha,0\right)}\left(\theta_C\right)V_{\alpha\beta}^mU_{C;\left(\gamma,\ell\right),\left(\beta,0\right)}^{\star}\left(\theta_C\right)$
\State \Return $\frac{1}{M} \sum_m \mathcal{C}_{\lambda\eta}\left(\rho^m,\ell_m\right)$
\EndFunction
\end{algorithmic}
\end{algorithm}
In practice, we optimize the circuit using the Limited Memory Broyden–Fletcher–Goldfarb–Shanno (LBFGS) algorithm implemented in PyTorch~\cite{NEURIPS2019_9015}.

\begin{figure*}[t]
  \begin{center}
\includegraphics[width=1.8\columnwidth]{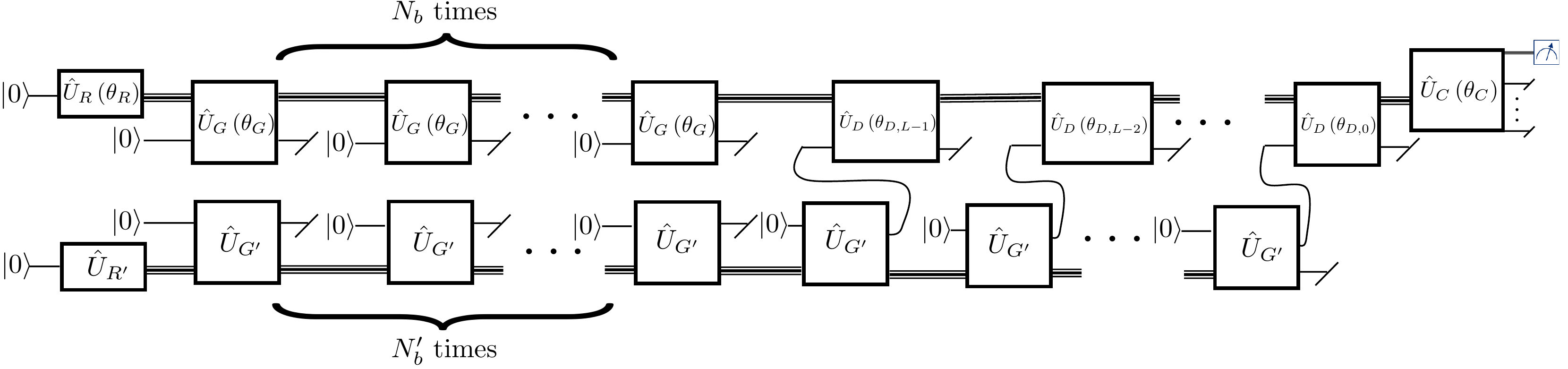}  
\caption{\label{fig:EI} \emph{Inference of an MPS with the discriminator architecture}.  The diagram shows inference of an iMPS test state (lower half of diagram) with unitaries $\hat{U}_{G'}$ and $\hat{U}_{R'}$ using the discrimination circuit with unitaries $\hat{U}_R$, $\hat{U}_G$, $\hat{U}_D$, and $\hat{U}_C$ (top half of diagram).  The test state generates its half-infinite density matrix using $N_b'$ iterations of the transfer operator encoded by $\hat{U}_{G'}$ applied to the boundary state generated by $\hat{U}_{R'}$, while the discriminator generates the feature space half infinite density matrix using $N_b$ iterations of $\hat{U}_G$ applied to the boundary state generated by $\hat{U}_{R}$ on a different bond register.  The test state generator then generates a single qubit of the test state at a time, this state is fed into the discriminator to condition its bond register, and the physical qubit is reset.  After $L$ qubits have been generated of the state, the test state bond qubits are discarded, the discriminator bond qubits are fed into the classifier circuit $\hat{U}_C$, and the measurement of the class register qubits defines the predicted class label $\ell$.}
\end{center}
\end{figure*}

\subsection{Model inference with entangled data}
\label{sec:EntInf}
Once the optimal parameters $\boldsymbol{\theta}_{\mathrm{opt}}$ for execution of the model without postselection have been determined, Algorithm \ref{alg:CFNoPS} can be simply adapted to inference the model by replacing the final cost function evaluation with sampling $\ell$ from $\rho_{\ell\ell}$.  As described, this algorithm applies to unentangled product state data, but this same procedure can be readily adapted to inference of entangled data.  Here, we present an algorithm for inferencing entangled data in the form of an iMPS model.  Namely, the test data is assumed to be provided in the form of unitaries $\hat{U}_{R'}$ and $\hat{U}_{G'}$ embedding a right boundary condition and a left-canonical MPS tensor $A$, respectively.  An algorithm to obtain this form for the ground state of a Hamiltonian with known matrix product operator (MPO) form is presented in Appendix \ref{app:iMPS}.  

As described in Sec.~\ref{sec:MPS}, the procedure to generate a finite state string from a left canonical iMPS is to first obtain the half-infinite density matrix by iteration of the transfer operator on the right boundary vector, and then to use the iMPS tensor with this density matrix as input to generate samples.  In practice, we will obtain the half-infinite density matrix using a finite number $N_b'$ of iterations as in Eq.~\eqref{eq:NbBurnIn}, with $N_b'$ a hyperparameter (set to $N_b$ in the results that follow).  Hence, the process for inferencing entangled state is (1) generate the half-infinite density matrix of the test state via $N_b'$ iterations of $\hat{U}_{G'}$ applied to the right boundary bond state generated by $\hat{U}_{R'}$ on a register of $\log_2\chi'$ bond qubits $|b'\rangle$; (2) generate the half-infinite density matrix of the feature space employed by the classifier by applying $N_b$ iterations of $\hat{U}_{G}\left(\theta_G\right)$ applied to the right boundary bond state generated by $\hat{U}_{R}\left(\theta_R\right)$ on a different register of $\log_2\chi$ bond qubits $|b\rangle$; (3) for $i=1,\dots,L$, generate a qubit of the test state using $\hat{U}_{G'}$ applied to the register $|b'\rangle$ and a single physical qubit in the $|0\rangle$ state, feed this state into the classifier by applying $\hat{U}_{D}\left(\theta_{D;L-i}\right)$ to the prepared state and the classifier bond register $|b\rangle$ and resetting the physical qubit; (4) apply the classifier unitary $\hat{U}_C\left(\theta_C\right)$ to the classifier bond register $|b\rangle$; (5) measure the subset of qubits defining the class register and discard all others.  This process is shown graphically in Fig.~\ref{fig:EI}.  The qubit resource requirements, provided the same physical qubit is utilized by both the test state and the discriminator to generate their half-infinite density matrices, are $\log_2\chi+\log_2\chi'+1$.  Utilizing two physical qubits--one for the test state and one for the discriminator--leads to the resource requirement $\log_2\chi+\log_2\chi'+2$, but enables the parallelized generation of the half-infinite density matrices shown in Fig.~\ref{fig:EI}.

\section{Demonstration on the transverse field Ising model}
\label{sec:Example}

In this section we describe an example application of the discriminator architecture, training, and test strategies put forth in Sec.~\ref{sec:TND}.  In particular, we consider the problem of discriminating the two phases of the spin-1/2 transverse field Ising model (TFIM) in one dimension, a canonical model for quantum phase transitions~\cite{sachdev2011quantum} that has been realized in a variety of physical systems~\cite{kim2011quantum,toskovic2016atomic,dutta_aeppli_chakrabarti_divakaran_rosenbaum_sen_2015}.  We demonstrate this application in both classical simulations and on Quantinuum's H1-2 trapped ion quantum computer.  The Hamiltonian for the antiferromagnetic TFIM is
\begin{align}
\label{eq:TFIM} \hat{H}_{\mathrm{TFIM}}&=\sum_{\langle i,j\rangle}\hat{\sigma}^z_i\hat{\sigma}^z_j-h\sum_i \hat{\sigma}^x_i\, ,
\end{align}
in which $\langle i,j\rangle$ denotes a sum over nearest-neighbor pairs $i,j$ and we have normalized units of energy to the antiferromagnetic coupling strength. For $h\gg 1$ the ground state is in a paramagnetic (PM) phase whose characteristics are captured by the classical state $|\rightarrow\dots\rightarrow\rangle$ with all spins pointing along the $x$ direction.  When $h\ll 1$ the ground state is in the antiferromagnetic (AFM) phase with properties similar to the classical states $|\uparrow\downarrow\dots\uparrow\downarrow\rangle$ and $|\downarrow\uparrow\dots\downarrow\uparrow\rangle$.  There is a quantum phase transition between the PM and AFM phases at $h=1$ in the thermodynamic limit, and we will train our discrimination architecture on data taken on either side of the transition with the goal of discriminating the phase of an unknown test quantum state.

\subsection{Description of dataset}
\label{sec:DS}

Our dataset will be comprised of single-shot measurements taken from the ground state of Eq.~\eqref{eq:TFIM} at some value of $h$.  We will define a single-shot measurement through a collection of $L$ Hermitian operators $\{\hat{O}_i\}, i=1,\dots,L$ which act on a subset of $L$ contiguous sites $\mathcal{L}$ from the infinite lattice indexed by $i$.  Each of these operators admits a spectral decomposition as $\hat{O}_i=\sum_{\lambda_i}\lambda_i|\lambda_i\rangle\langle \lambda_i|$.  A single-shot measurement is now defined as a simultaneous projective measurement of all of the operators $\{\hat{O}_i\}$, which we can denote as a collection of the associated eigenvalues $\{\lambda_i\}$.  Associated with this measurement is a classical product state (CPS) $|\lambda_1\dots\lambda_L\rangle$, which is the state produced by the projective measurement.  The association is provided by the fact that the measurement $\{\lambda_i\}$ from a state $|\psi\rangle$ occurs with probability $\mathrm{Tr}\left[|\lambda_1\dots \lambda_L\rangle\langle \lambda_1\dots \lambda_L| \mathrm{Tr}_{i\notin \mathcal{L}}|\psi\rangle\langle\psi|\right]$.  A dataset is comprised of a collection of these CPSs, randomly chosen from the ground state $|\psi\rangle$ at specific points within the phase diagram, together with labels $\ell$ denoting the quantum phase of the state at that point, $\mathcal{T}=\{|\boldsymbol{\lambda}_m\rangle,\ell_m,m=1,\dots,M$\}.

We generate our single-shot measurement training data using MPS methods.  Namely, we find the ground state at transverse field strength $h$ on $F\gg L$ sites using standard variational MPS methods~\cite{schollwock2011density}, and then sample CPSs from this MPS given an operator basis $\{\hat{O}_i,i\in\mathcal{L}\}$ on a subset of $L$ sites $\mathcal{L}$.  In the results presented in this work, we take $F=32$, a truncated weight of $\varepsilon=10^{-6}$, maximum bond dimension of $\chi_{\mathrm{max}}=40$, $L=6$, and $\mathcal{L}$ to be the $L$ sites in the center of the chain.  We take half of our samples in the uniform $z$ basis $\{\hat{O}_i=\hat{\sigma}^z_i, \forall i\in \mathcal{L}\}$ and half in the uniform $x$ basis $\{\hat{O}_i=\hat{\sigma}^x_i, \forall i\in \mathcal{L}\}$.  These choices reflect how we envision our discrimination architecture could be used in a near-term setting; the central region of a finite-size experimental system will be measured and taken to representative of the bulk of an infinite system.  Present-day quantum simulators, e.g. those based on trapped cold atoms, do not universally feature single-site addressability, but many have the ability to change the global measurement basis, e.g. by applying a uniform microwave pulse.  Generating uncorrelated CPS samples from an MPS has been described before in the context of minimally entangled typical thermal state (METTS) algorithms, and is reviewed in Appendix \ref{app:CPSSampling}.

We take the states defining the PM and AFM phases to have $h=10$ and $h=0.1$, respectively, and generate 1000 shots at each value of $h$ in both the $x$ and $z$ basis.  These shots are randomly reshuffled using an $80/20$ split into a training set and a test set.  We note that the ground state in the AFM phase is doubly degenerate, with the ground states being distinguished by the global spin flip operator $\hat{X}=\prod_{i=1}^{F}\hat{\sigma}^x_i$ that generates a $\mathbb{Z}_2$ symmetry.  Numerical noise in the initialization of the MPS ground state search procedure can result in finding a symmetry broken state in this manifold.  If all training measurements were taken from a single MPS representation, this would result in significant bias compared to measurements taken from an experiment, in which a different symmetry broken state could be obtained for each realization.  In order to avoid this phenomenon, we add a term $-h_{\mathbb{Z}_2} \hat{X}$ to the Hamiltonian with $h_{\mathbb{Z}_2}\sim 10$ to ensure that the $\mathbb{Z}_2$-symmetric ground state is chosen and there is no symmetry-breaking bias in the training data.

Finally, for entangled inferencing of a model trained on the product state data just described (see Sec.~\ref{sec:EntInf}), we also generate iMPS models of the ground state with small bond dimension $\chi\ll \chi_{\mathrm{max}}$ through a novel optimization procedure described in Appendix \ref{app:iMPS}.  This algorithm produces a translationally invariant iMPS tensor $A'$ in left-canonical form together with a boundary bond state $R'$ that optimizes the production of the true half-infinite density matrix obtained from $|\psi\left[A'\right]\rangle$ from a finite ``burn-in" procedure with a fixed number $N_b'$ of iterations.  Following optimization, these tensors can be gauged and compiled as described in Sec.~\ref{sec:classtraining} to define circuits for inferencing as described in Sec.~\ref{sec:EntInf}.

\subsection{Training and evaluation}
\label{sec:ClassTandE}
Here, we apply the classical training and compilation procedures of Sec.~\ref{sec:classtraining} to the dataset described in Sec.~\ref{sec:DS}.  Our main metric of performance will be the $F_1$ score, defined from precision $p$ and recall $r$ as 
\begin{align}
p_i &=\frac{C_{ii}}{\sum_j C_{ji}}\, ,\\
r_i &=\frac{C_{ii}}{\sum_j C_{ij}}\, ,\\
\left[F_1\right]_i&=2\frac{p_ir_i}{p_i+r_i}\, .
\end{align}
In these expressions, $\mathbb{C}$ is the confusion matrix indexed by classes whose element $C_{ij}$ represents the number of data elements predicted to be in class $j$ whose truth class is $i$.  

As a classical machine learning benchmark, we train a random forest classifier using the implementation in scikit-learn~\cite{scikit-learn} with the hyperparameter \verb#n_estimators# set to 20.  Here, the classical product state single-shot measurements are flattened into classical vectors of dimension $2L$ as
\begin{align}
|\mathbf{x}\rangle&=\left(\sum_{i_0} x_{0,i_0} |i_0\rangle\right)\otimes \dots \otimes \left(\sum_{i_{L-1}} x_{L-1,i_{L-1}} |i_{L-1}\rangle\right)\\
&\to \left(\begin{array}{ccccc} x_{0,0}&x_{0,1}&\dots &x_{L-1,0}&x_{L-1,1}\end{array}\right)^T\, ,
\end{align}
and these classical representations are used for learning.  This flattened representation can capture superpositions at the single qubit level, but not entanglement between qubits.  We note that no claims are made about the optimality of this classical data encoding or classifier, we are merely using it as a reasonable classical benchmark.

Following training of the tensor elements using manifold gradient descent on the cost function in algorithm \ref{alg:CFtensor}, we find class-averaged $F_1$ scores of $\bar{F}_1=0.814$, 0.910, and 0.939 on the training data  for models with $\chi=2$, 4, and 8, corresponding to 1, 2, and 3 bond qubits, respectively.  The classical random forest benchmark achieves a class-averaged $F_1$ score of 0.973 on the training set.  We then compile the associated unitary embeddings using the greedy heuristics described in Sec.~\ref{sec:classtraining} with a squared two-norm tolerance of $4\times 10^{-4}$, a basis of CNOT entangling gates and $R_y$ single-qubit rotations, and without any restrictions on the qubit pairs CNOTs can be applied to.  The CNOT counts for the various compiled unitaries are collected in Table~\ref{tab:CNOTcounts}.

\begin{table}[h]
\begin{tabular}{|c|c|c|c|}
\hline &$\chi=2$&$\chi=4$&$\chi=8$\\
\hline $\hat{U}_R$&0&1&3\\
\hline $\hat{U}_G$&2&9&42\\
\hline $\hat{U}_D$&2&9&41\\
\hline $\hat{U}_C$&2&3&8\\ \hline
\end{tabular}
\caption{\label{tab:CNOTcounts} CNOT counts for the compiled unitaries with increasing bond dimension.}
\end{table}

\begin{figure}[t]
  \begin{center}
\includegraphics[width=0.93\columnwidth]{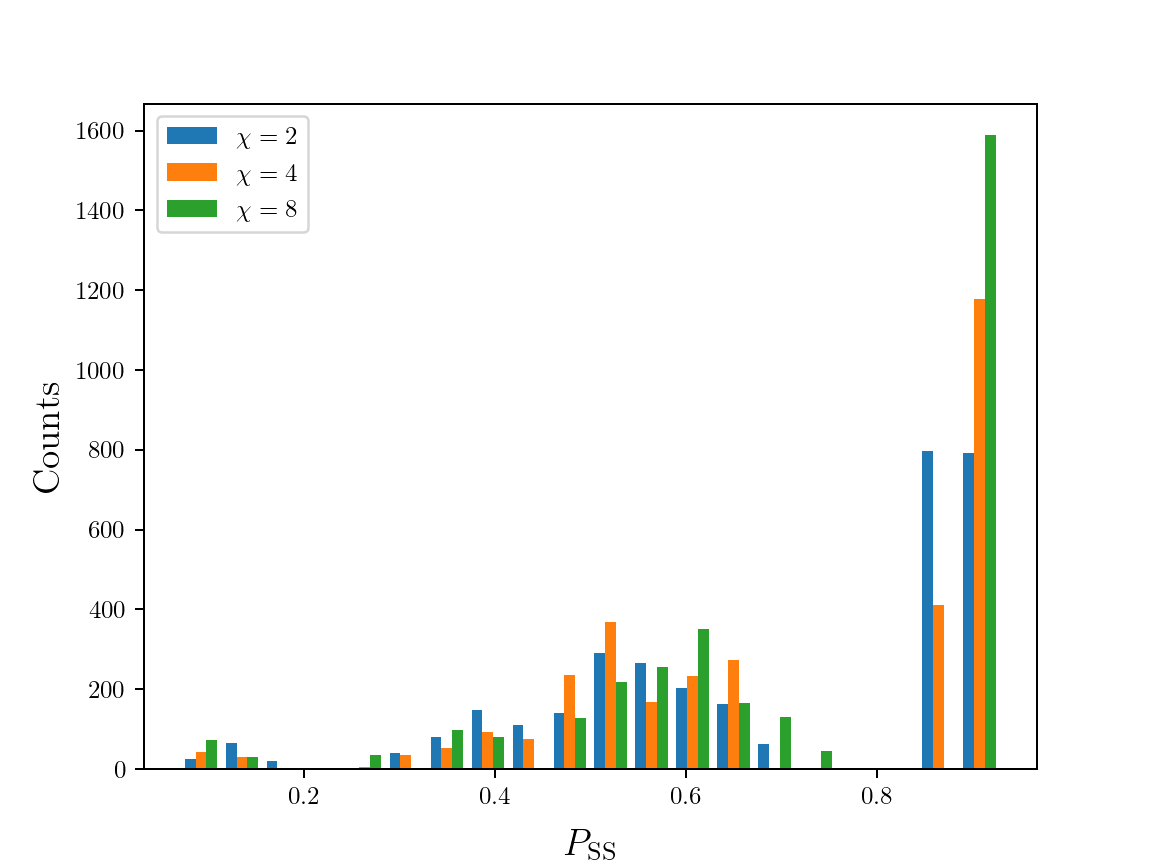}  
\caption{\label{fig:PSS} \emph{Dependence of single-shot probability of success on bond dimension}.  Histograms of the single-shot probability of correct classification $P_{\mathrm{SS}}$ over the training set are shown for $\chi=2,4,8$.  As the bond dimension increases, the average probability of correct classification also increases.}
\end{center}
\end{figure}

While the finite tolerance of compiling the unitaries does not significantly impact the model when run with postselection, the $\bar{F}_1$ scores using these parameters without postselection are 0.360, 0.364, and 0.228.  Following 30 epochs of LBFGS optimization of the compiled unitary parameters with the cost function hyperparameters $\lambda=0.9$, $\eta=2.0$, we find optimized $\bar{F}_1$ scores of 0.764, 0.822, and 0.860.  In addition to the $\bar{F}_1$ scores, an important metric as $\chi$ increases is the single-shot probability of success $P_{\mathrm{SS}}=\rho_{\ell_m,\ell_m}$, which determines the frequency with which the classifier will return the correct label in a single execution of the circuit.  As shown in Fig.~\ref{fig:PSS}, increasing the bond dimension not only increases the average number of correct classifications (captured by the $\bar{F}_1$ score), but also the probability that a single run of the circuit produces the correct label, on average.  We note that the clustering of highest probabilities near $0.9$ is influenced by the choice of the hyperparameter $\lambda$; different choices of hyperparameters, possibly bond-dimension-dependent, can alter this distribution.

\begin{figure}[t]
  \begin{center}
\includegraphics[width=0.93\columnwidth]{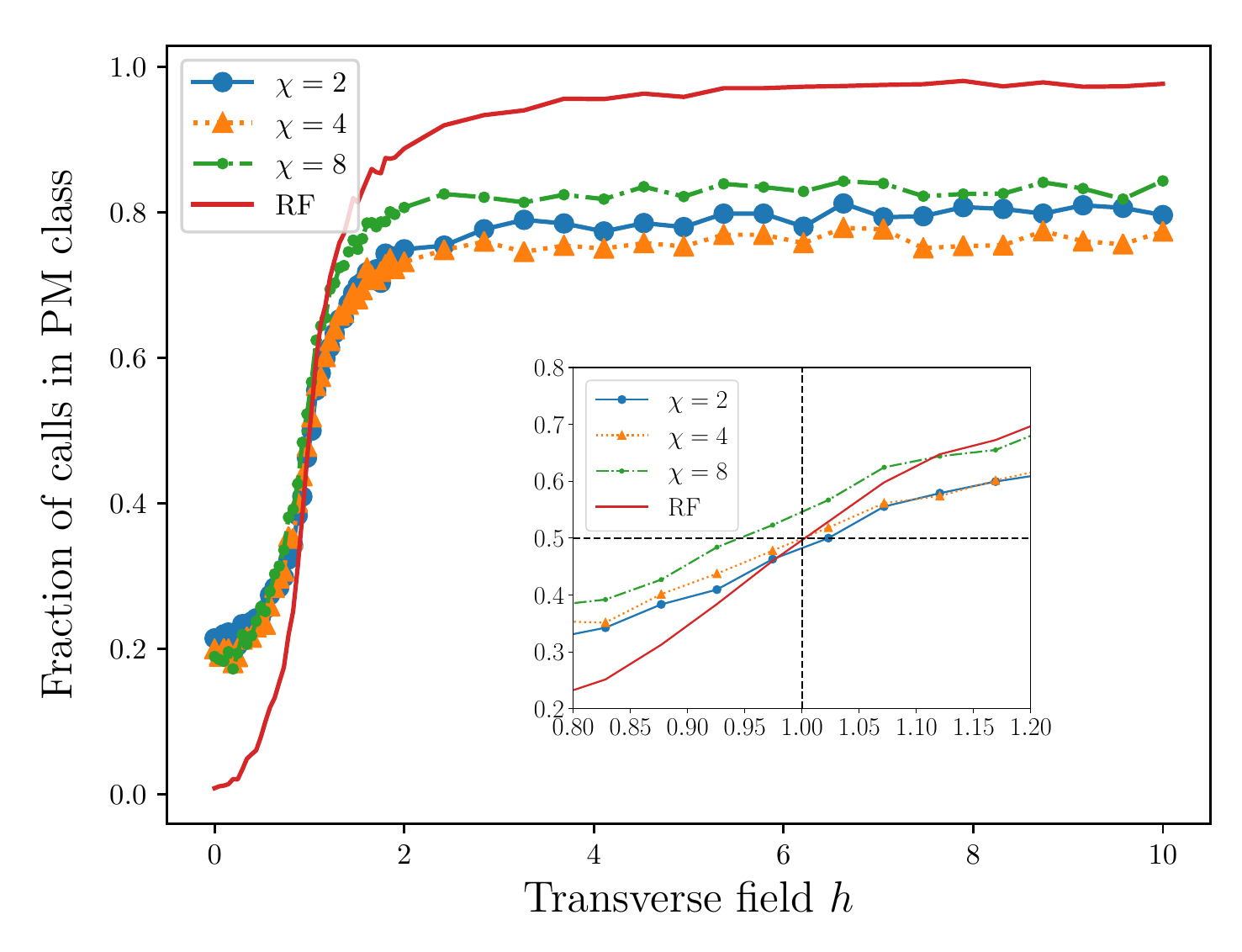}  
\caption{\label{fig:PDFig} \emph{Predicted phases from the tensor network discriminator and classical benchmark; product state input data}.  The probability that a classifier predicts the PM class, averaged over the test data, is shown as a function of the transverse field $h$.  The inset shows a zoom around the true phase transition point, $h=1$, demonstrating that all classifiers have a bias of a few percent.}
\end{center}
\end{figure}

A canonical use case for a quantum classifier is determining the quantum phase of the state producing the measurements, where the classifier has been trained on data deep in the phases to be distinguished.  In this use case, a key metric is the distance between the phase transition point predicted by the classifier and the true transition point.  In the case of binary classification, the predicted phase transition point can be taken as the place where the classifier predicts $50\%$ probability for both phases.  In order to test this use case, we produced ground state representations in the range $h\in\left[0,10\right]$ using MPS variational ground state search and sampled 1000 single-shot measurements in the $x$ and $z$ basis for each $h$, following the procedure described in Sec.~\ref{sec:DS}.  The fraction of data predicted to be in the PM class for the $\chi=2,4,8$ tensor network discriminators and the random forest benchmark are shown in Fig.~\ref{fig:PDFig}.  We see slight improvements in the probability of correct classification as the bond dimension is increased, in accordance with the behavior seen in the training set in Fig.~\ref{fig:PSS}, while the random forest classifier shows the best performance by this metric.  As seen in the inset, all classifiers are remarkably unbiased in their determination of the phase transition point at $h=1$ even though they were trained on data at $h=0.1$ and $h=10$; all classifiers are within a few percent of this value.

\begin{figure}[t]
  \begin{center}
\includegraphics[width=0.93\columnwidth]{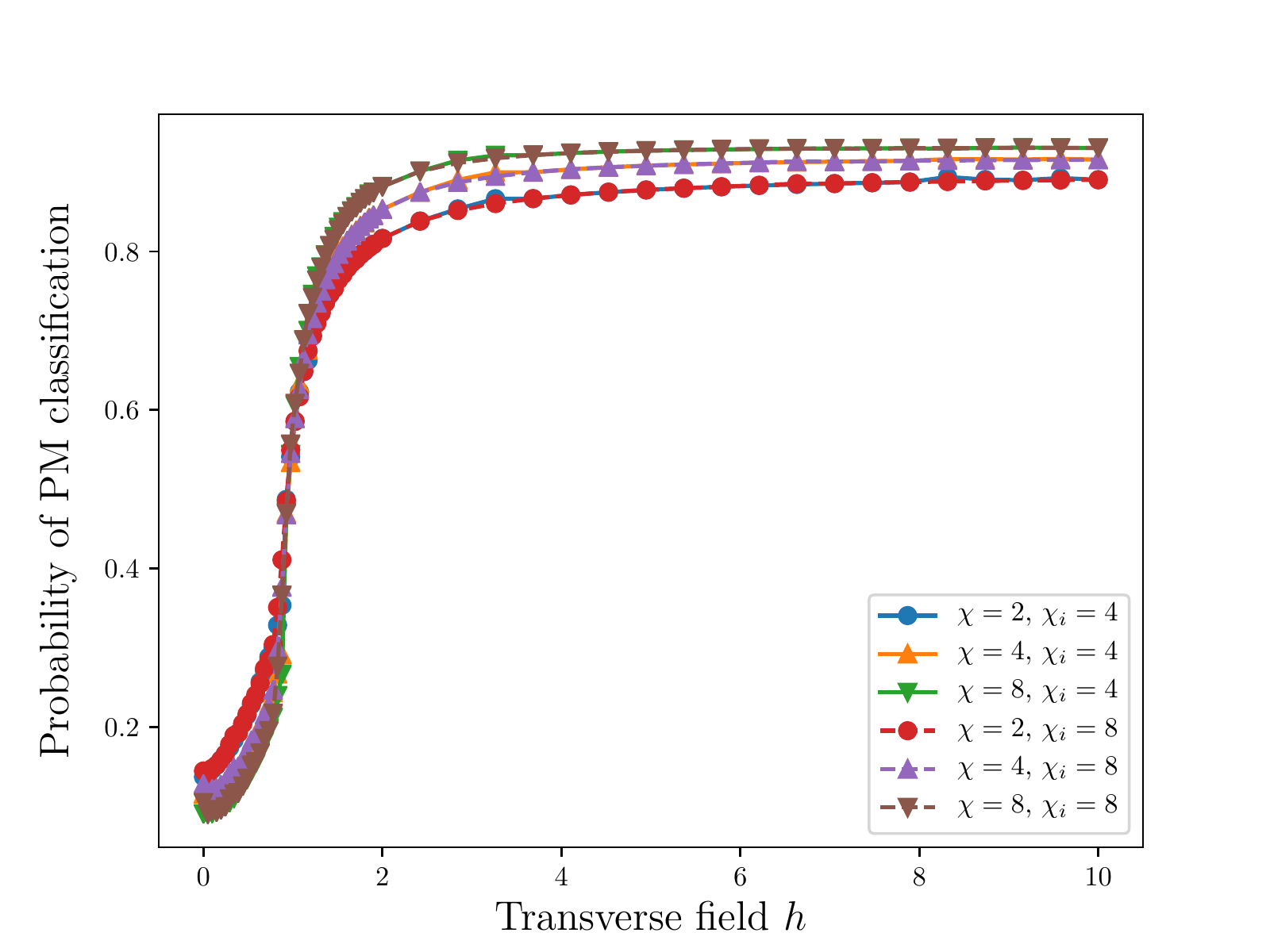}  
\caption{\label{fig:EPDFig} \emph{Predicted phases from the tensor network discriminator and classical benchmark; entangled input data}.  The probability of a classifier returning a call in the PM class for a single run of the circuit is shown as a function of the transverse field $h$.  As the bond dimension of the classifier increases, the probability of correct classification increases.  The performance of all classifiers is relatively unchanged as the bond dimension of the iMPS generating the test data is increased from $\chi_i=4$ to $\chi_i=8$.}
\end{center}
\end{figure}

The final use case we consider is inferencing of our models on entangled input data.  To generate the test data for this use case, we produce iMPS models for the ground state at a variety of transverse field strengths $h$ as described in Appendix \ref{app:iMPS}.  We stress that this process is different than the finite-size ground state search used to generate the single-shot measurement training and test data.  We then inference our tensor network models using a string of $L=6$ qubits of entangled data generated from these iMPS models as described in Sec.~\ref{sec:EntInf}.  The results are shown in Fig.~\ref{fig:EPDFig}.  Here, we plot the probability that the classifier determines the input state to be in the PM class in a single run of the circuit, as opposed to the case of Fig.~\ref{fig:PDFig}, in which probabilities were inferred by averaging the 2000 elements of the test set.  Increasing the bond dimension of the classifiers improves their performance in the sense that the single-run probability of correct classification increases.  We see little dependence of the performance on the bond dimension of the input data, $\chi_i$.  This process of entangled inference following training of the classifier on single-shot measurements provides a metric for determining the consistency of experimental data (single-shot measurements) with a reduced-order model (entangled input data according to a particular model Hamiltonian).  Finally, we note that there is no classical benchmark here, as the inferencing of entangled data is an inherently quantum process.  It is interesting to note that the performance of the classifiers on entangled input data even at the smallest bond dimension is higher than the classifiers evaluated on single-shot data at any of the evaluated bond dimensions, even though the classifiers were trained only on single-shot measurements.

\subsection{Hardware implementation}

\begin{figure}[t]
  \begin{center}
\includegraphics[width=0.93\columnwidth]{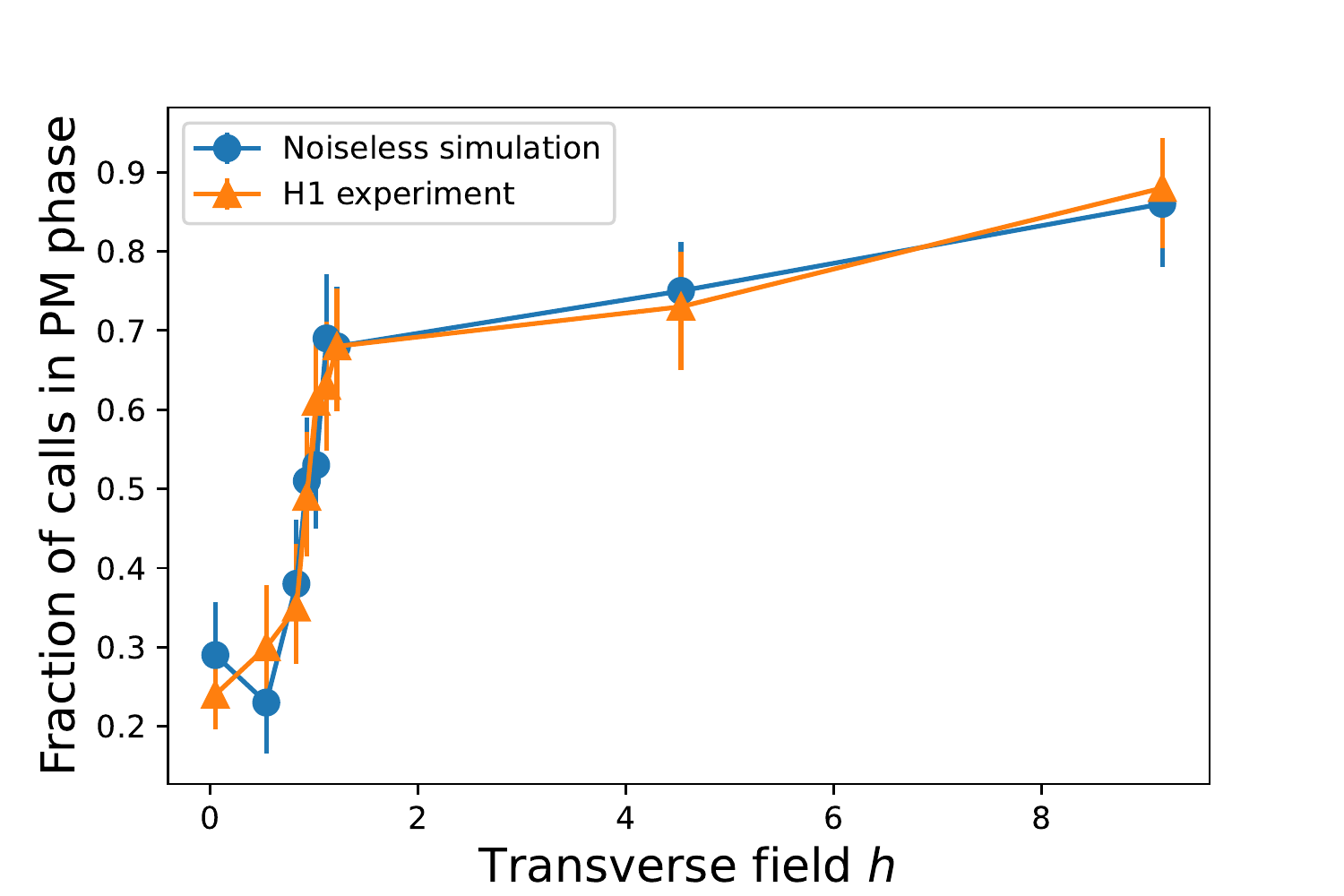}  
\includegraphics[width=0.93\columnwidth]{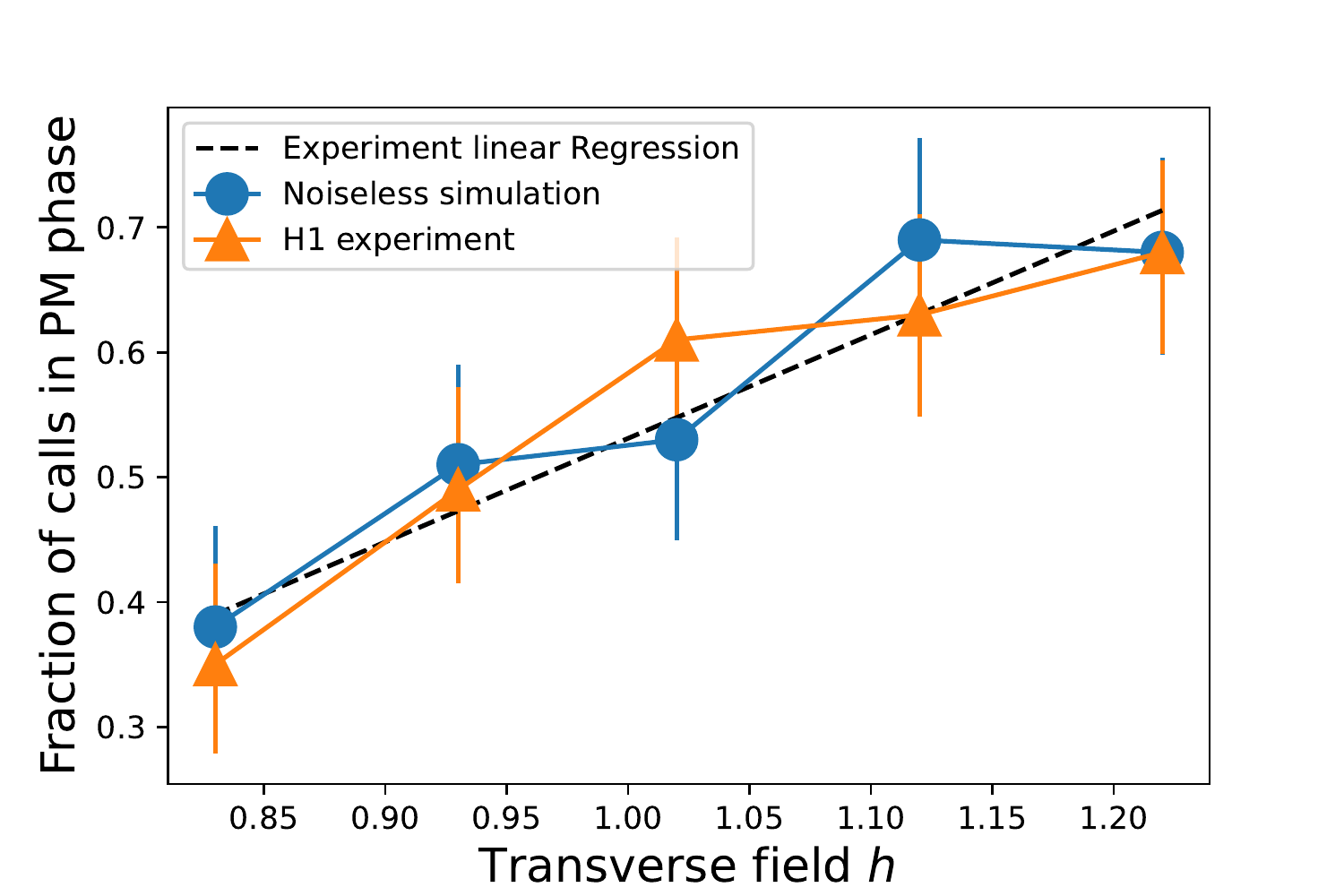}  
\caption{\label{fig:H1chi2}\emph{Performance of $\chi=2$ model on entangled data.}  The top panel shows the predictions of the $\chi=2$ tensor network discriminator evaluated on entangled iMPS input data with $\chi_i=4$, with the triangles corresponding to the mean of experimental results on the H1 machine and the circles corresponding to simulations of the circuit execution in the absence of hardware noise.  All points correspond to 100 shots, and error bars are 90\% Wilson confidence intervals.  The bottom panel is a zoom around the transition point, together with the linear regression model predicting a transition at $h=$ 0.962.}
\end{center}
\end{figure}

\begin{figure}[t]
  \begin{center}
\includegraphics[width=0.93\columnwidth]{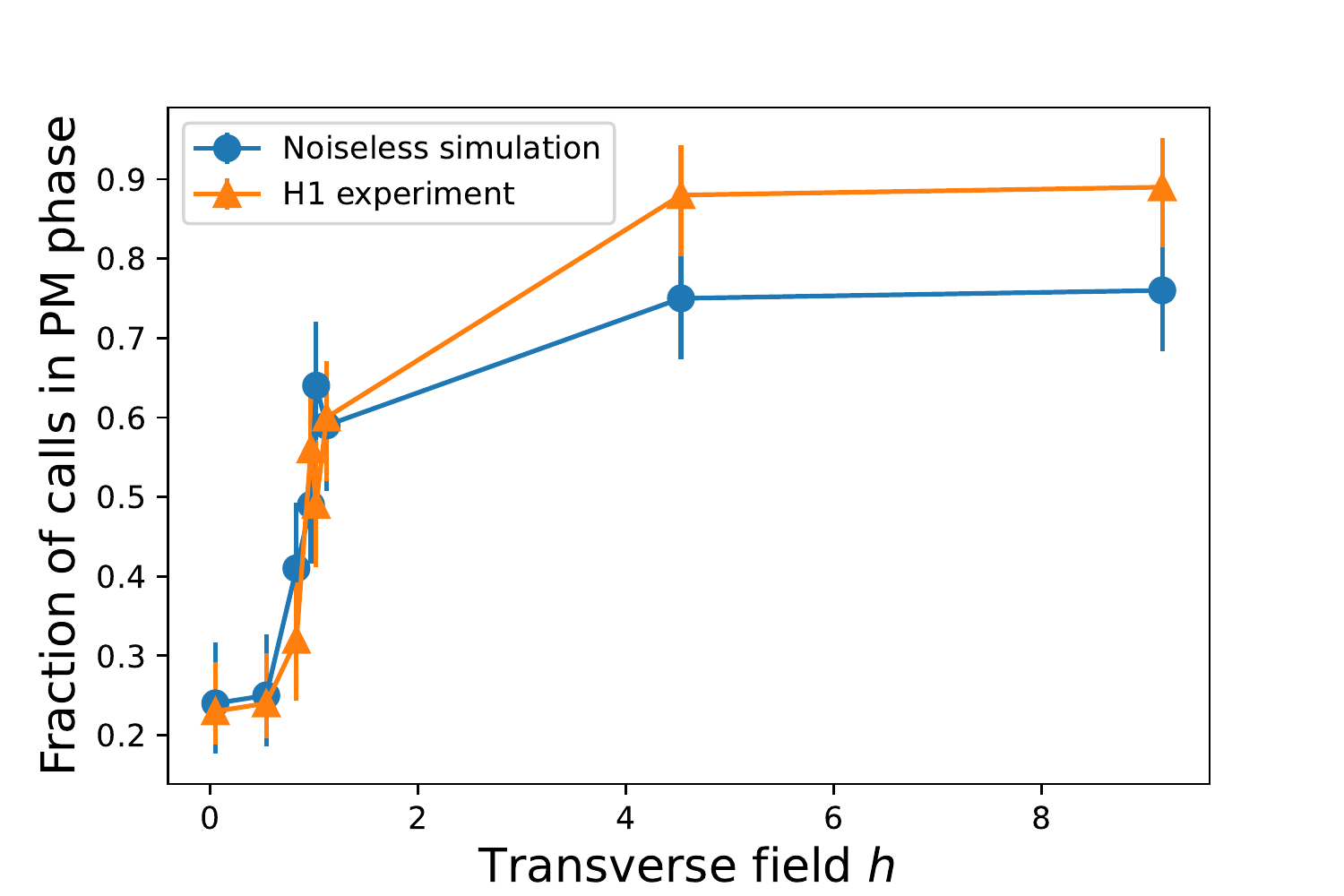}  
\includegraphics[width=0.93\columnwidth]{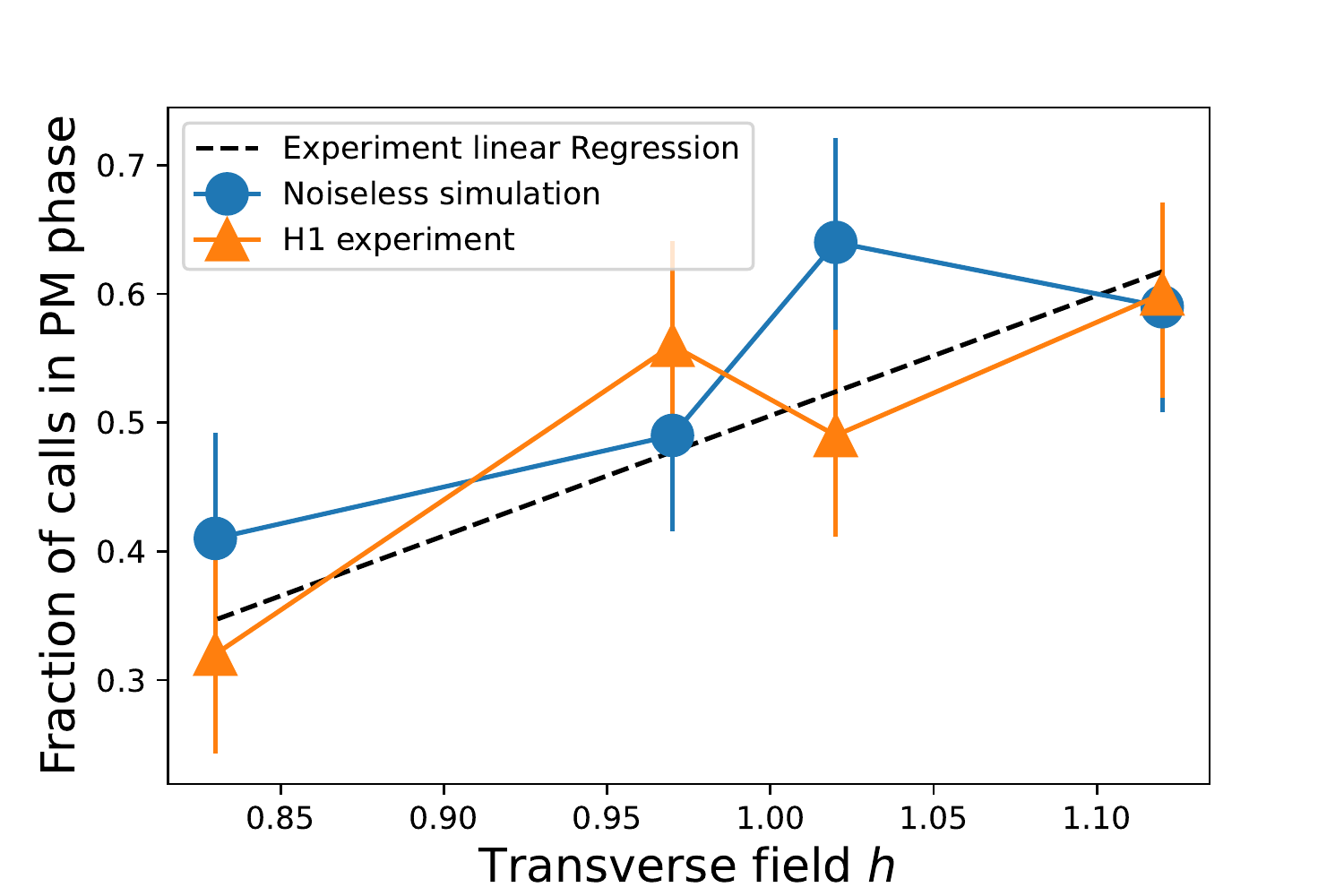}  
\caption{\label{fig:H1chi4}\emph{Performance of $\chi=4$ model on entangled data.}  The top panel shows the predictions of the $\chi=4$ tensor network discriminator evaluated on entangled iMPS input data with $\chi_i=4$, with the triangles corresponding to the mean of experimental results on the H1 machine and the circles corresponding to simulations of the circuit execution in the absence of hardware noise.  All points correspond to 100 shots, and error bars are 90\% Wilson confidence intervals.  The bottom panel is a zoom around the transition point, together with the linear regression model predicting a transition at $h=$ 0.994.}
\end{center}
\end{figure}

We now turn to results obtained on Quantinuum's H1-2 trapped ion quantum computer~\cite{suppl}.  A key consideration when running the model on quantum hardware is the number of runs of the circuit required to obtain a high-confidence result for the predicted class.  Each run of the circuit with data instance $|\mathbf{x}\rangle$ returns a class label $\ell$ which is 0 with an unknown probability $p_0\left(\mathbf{x}\right)$.  The task of the classical postprocessing is to infer what $p_0\left(\mathbf{x}\right)$ is from a finite number $n$ of runs of the circuit, and assign the class of $\mathbf{x}$ as $0$ if $p_0\left(\mathbf{x}\right)>0.5$ and 1 if $p_0\left(\mathbf{x}\right)<0.5$.  In the ideal case where hardware noise is independent and identically distributed, each run of the circuit constitutes a Bernoulli trial and we can utilize binomial confidence intervals to estimate the uncertainty in the model's predictions.  In what follows, we will employ the Wilson score interval, which utilizes a normal approximation to the binomial to derive closed-form expressions for the confidence interval in terms of $z$, the $1-\frac{\alpha}{2}$ quantile of the normal distribution with target error rate $\alpha$.  Namely, the Wilson score interval predicts
\begin{align}
p_0\left(\mathbf{x}\right)&\approx \frac{n_0+z^2/2}{n+z^2}\pm \frac{z}{n+z^2}\sqrt{\frac{n_0n_1}{n}+z^2/4}\, ,
\end{align}
where we have suppressed the dependence of $n_0$, $n_1$, and $n$ on $\mathbf{x}$ on the right hand side to keep the notation uncluttered.  In the results presented, we use the 90\% confidence interval, corresponding to $z=1.645$.  A complementary online algorithm which determines when a sufficient number of shots has been taken according to a Bayesian credible interval is given in Appendix ~\ref{app:Bayes}.

Given the better performance of the models inferenced on entangled input data vs.~single-shot data in classical simulation (see Sec.~\ref{sec:ClassTandE}), we utilized the entangled inference strategy on the quantum hardware.  Also noting the relative insensitivity of our model performance to the bond dimension of the entangled input data (see Fig.~\ref{fig:EPDFig}), we utilized a fixed input data bond dimension of $\chi_i=4$ for all experiments.  For each value of the transverse field $h$ considered, the circuit was run for 100 shots, and the Wilson score interval utilized to obtain the uncertainty due to a finite number of shots.  In order to identify the predicted transition point, we take the points closest to 50\% predicted probability in the PM phase and fit them to a linear regression model with the Wilson score intervals as errors.  Figs.~\ref{fig:H1chi2} and \ref{fig:H1chi4} show the results for $\chi=2$ and $4$, respectively.  In both figures, the orange triangles correspond to the fraction of shots classifying the input state as being in the PM class, the blue circles are the same quantity computed using an H1 simulator in the absence of hardware noise, the error bars are the 90\% Wilson confidence intervals, and the black dashed line is the prediction of the fit linear regression model.  Identifying the point of ``maximal confusion" of the classifier as the transition point, the linear regression predicts $h=0.962$ and $0.994$ for the $\chi=2$ and $\chi=4$ models, respectively.  The mean absolute error (MAE) and $R^2$ coefficient of determination of the two linear models over the data used in the regression are $(0.03,0.90)$ and $(0.04,0.80)$ for $\chi=2$ and $4$, respectively, indicating that the lower bond dimension data is slightly better fit and the linear model captures more of the variance in that data.  This is to be expected on account of the deeper circuits utilized for the $\chi=4$ model.

\section{Conclusions and outlook}
\label{sec:Concl}

We presented a workflow for classifying a translationally invariant quantum state based on $L$ qubits of quantum data extracted from it using a tensor network (TN) architecture with a one-dimensional matrix product state (MPS) topology.  In our discriminator architecture, a register of $\log_2\chi$ ancilla qubits forms an effective bond space of dimension $\chi$ mediating correlations between physical degrees of freedom.  Similar to encoder methods in classical machine learning, this bond space stores an effective reduced-dimensional latent representation of the data the model represents.  Inferencing of our model begins with preparing the ancilla qubits in an initial state which encodes a prior distribution over a collection of latent quantum feature vectors.  A series of quantum operations coupling these ancilla qubits with a register of physical qubits then conditions the state of the ancillae on the input quantum data.  A final quantum operation predicts the class of the input data depending on the state of the ancilla qubits and stores the result in a subset of the ancillae.  We presented methods for robust and efficient classical preconditioning of this discriminator architecture from a dataset of single-shot experimental measurements, consisting of optimization of the isometric tensors in a classical tensor network representation using manifold optimization techniques, compilation of the isometric tensors into quantum operations using hardware-aware greedy compilation heuristics, and further optimization of the parameters of the resulting compiled quantum circuit to remove the post-selection requirements of the isometric tensor network model.  Following training, the resulting model can be inferenced with quantum data on quantum hardware.  Our architecture enables the discovery of model circuits in a data-driven fashion without specifying the gate depth or composition ahead of time, and aids in trainability by enabling efficient classical preconditioning and requiring measurements of only a small subset of model qubits.

We demonstrated the training of our TN discriminator on a synthetic dataset of data extracted from the one-dimensional transverse field Ising model (TFIM) in its paramagnetic (PM) and antiferromagnetic (AFM) phases.  Following training on single-shot experimental data, we analyzed the performance of the model inferenced on product state or entangled input data.  We found that increasing the bond dimension of the model, amounting to adding more qubits to the ancilla register in the quantum representation, improved both the average classification accuracy and the probability of correct classification in a single run of the circuit, and that inference of entangled data had better average performance than inference of unentangled data.  Using TN models of bond dimension $\chi=2$ and $4$ trained on single-shot data deep in the PM and AFM phases, we ran experiments on Quantinuum's model H1-2 trapped ion quantum computer inferencing these models against entangled input data modeled as MPSs of bond dimension $\chi_i=4$ across the phase diagram.  Using linear regression near the point of maximal confusion of our models (i.e., 50\% prediction in each phase), we extracted predictions of the phase transition at $h=0.962$ and $0.994$ for bond dimension $\chi=2$ and $\chi=4$, comparing well with the known location of $h=1$.

Our results demonstrate that classical techniques for optimizing TNs can be utilized to define and optimize well-performing machine learning models for quantum hardware in a data-driven and hardware-aware fashion.  By analyzing a series of classically pre-conditioned model architectures of increasing bond dimension, model architectures can be scaled towards the classically intractable regime in which quantum models may display an advantage in expressibility.  While we focused on the case of MPS models, the generalization to other TN topologies is straightforward.  In addition, looking beyond machine learning, many of the tools we have developed can be brought to bear on efficiently preconditioning and scaling quantum TN models for other use cases, such as quantum simulation and error correction~\cite{PhysRevLett.113.030501,Jahn_2021}.

\section{Acknowledgements}

We would like to thank Matt Abernathy and Giuseppe D'Aguanno for collaboration on related work.  M.L.W., P.T., and G.Q. would like to acknowledge funding from the Internal Research and Development program of the Johns Hopkins University Applied Physics Laboratory.  K.R.A.H acknowledges support from the Welch Foundation through Grant No. C1872, the Office of Naval Research Grant No. N00014-20-1-2695, and the National Science Foundation through Grants Nos. PHY1848304 and CMMI2037545.  K.R.A.H.~also  benefited from discussions at the KITP, which was supported in part by the National Science Foundation under Grant No.~NSF PHY-1748958.

\appendix

\section{Algorithm for optimizing an iMPS in left-canonical form}
\label{app:iMPS}
In this appendix we provide an algorithm for obtaining a quantum circuit for the iMPS description of the ground state of a Hamiltonian $\hat{H}$ with a known matrix product operator (MPO) form.  While the iMPS algorithm put forth by McCulloch~\cite{mcculloch2008infinite} (building on the seminal iDMRG algorithm of White~\cite{white1992density}) works well for this purpose, the algorithms herein provide a simple and robust means for obtaining iMPS representations directly in left-canonical form, from which our quantum compilation methods can be applied to generate low-depth quantum circuits.

We define the action of the generalized left transfer operator on a $\chi\times \chi$ matrix $\mathbb{E}$ as
\begin{align}
{T}_{L,\hat{O}}\left(A; \mathbb{E}\right)&=\sum_{i,i'}\langle i|\hat{O}|i'\rangle \mathbb{A}^{i\dagger} \mathbb{E} \mathbb{A}^{i'}\, ,
\end{align}
where $\hat{O}$ is an operator acting on the single-site Hilbert space and $A$ is the translationally invariant iMPS tensor.  Hence, the left transfer operator of the identity for an iMPS in left-canonical form has the $\chi\times\chi$ identity matrix $\mathbb{I}$ as an eigenmatrix with eigenvalue 1, ${T}_{L,\hat{I}}\left(A; \mathbb{I}\right)=\mathbb{I}$.  Additionally, defining the corresponding right generalized transfer matrix
\begin{align}
{T}_{R,\hat{O}}\left(A; \mathbb{E}\right)&=\sum_{i,i'}\langle i|\hat{O}|i'\rangle \mathbb{A}^{i} \mathbb{E} \mathbb{A}^{i'\dagger}\, ,
\end{align}
we see that the right transfer operator of the identity admits the half-infinite reduced density matrix as an eigenmatrix with eigenvalue 1 in the case of a left-canonical iMPS, ${T}_{L,\hat{I}}\left(A; \rho_{\mathrm{half-infinite}}\right)=\rho_{\mathrm{half-infinite}}$.  Defining a matrix representation of the transfer operator as
\begin{align}
\left[T_{\hat{O}}\left(A\right)\right]_{\left(\alpha,\alpha'\right),\left(\beta\beta'\right)}&=\sum_{ii'} A^{i\star}_{\alpha\beta}\langle i|\hat{O}|i'\rangle A^{i'}_{\alpha'\beta'}\, ,
\end{align}
we see that the left and right transfer operators are just the left- and right- action of this transfer operator, and so $T_{\hat{I}}\left(A\right)$ admits $\mathbb{I}$ and $\rho_{\mathrm{half-infinite}}$ as left- and right-eigenmatrices with eigenvalue 1, respectively.

The transfer operator formalism is useful to extract fixed-point relations of the iMPS state, such as the eigenmatrix relations described in the last paragraph.  For a Hamiltonian with MPO representation defined by a translationally invariant MPO tensor $W_{\alpha\beta}^{ii'}$ that is lower triangular in the bond space, this same formalism will enable us to extract the energy density of an iMPS in the thermodynamic limit, as described in Ref.~\cite{mcculloch2008infinite,michel2010schur}.  Several works, e.g. Refs.~\cite{mcculloch2007density}, have discussed how to put common Hamiltonian terms into this lower triangular ``canonical form" and compose them.  For concreteness, we will consider the MPO representation of the TFIM used in the main text, given as
\begin{align}
\label{eq:TFIMMPO}\hat{W}&=\left(\begin{array}{ccc} \hat{I}&0&0\\ \hat{\sigma}_z&0&0\\ -h\hat{\sigma}^x&\hat{\sigma}_z&\hat{I}\end{array}\right)\, .
\end{align}
For each value of the bond index of the MPO matrix we can associate a matrix $\mathbb{E}_{\alpha}$, and the fixed-point relations for these matrices associated with the left action of generalized transfer operators take the form
\begin{align}
\mathbb{E}_{\alpha}&=T_{L\hat{W}_{\alpha\alpha}}\left(A,\mathbb{E}_{\alpha}\right)+\sum_{\beta>\alpha}T_{L\hat{W}_{\beta\alpha}}\left(A,\mathbb{E}_{\beta}\right)\, .
\end{align}
Because of the lower triangular nature of the MPO, we can solve for the terms in order, starting from the last ($\alpha=\chi_W$, with $\chi_W$ the bond dimension of the MPO).  Namely, we find the fixed-point relation
\begin{align}
\mathbb{E}_{\chi_W}&=T_{L\hat{I}}\left(A,\mathbb{E}_{\chi_W}\right)\, ,
\end{align}
and so $\mathbb{E}_{\chi_W}=\mathbb{I}$ for an iMPS in left-canonical form.  For the case of the TFIM MPO, Eq.~\eqref{eq:TFIMMPO}, the next matrix is
\begin{align}
\mathbb{E}_{2}&=T_{L\hat{\sigma}_z}\left(A,\mathbb{E}_{\chi_W}\right)\, ,\\
&=T_{L\hat{\sigma}_z}\left(A,\mathbb{I}\right)\, .
\end{align}
Hence, this is not a system of equations but a definition.  The one remaining fixed-point relation is
\begin{align}
&\mathbb{E}_1=T_{L\hat{I}}\left(A,\mathbb{E}_{1}\right)+T_{L\hat{\sigma}_z}\left(A,\mathbb{E}_{2}\right)-hT_{L\hat{\sigma}_x}\left(A,\mathbb{E}_{\chi_W}\right)\, ,\\
&\Rightarrow \mathbb{E}_1\left(\mathbb{I}-T_{\hat{I}}\left(A\right)\right)=T_{L\hat{\sigma}_z}\left(A,T_{L\hat{\sigma}_z}\left(A,\mathbb{I}\right)\right)-hT_{L\hat{\sigma}_x}\left(A,\mathbb{I}\right)\, .
\end{align}
Recalling that the transfer operator $T_{\hat{I}}\left(A\right)$ has the identity as a left eigenmatrix, we see that the left-hand side of this (left)- system of equations is singular, meaning that we can add any multiple of the identity to a solution and obtain another solution.  We can find the solution with zero weight on the identity by removing the components in this direction from the right hand side matrix.  This is achieved by recalling that $\rho_{\mathrm{half-infinite}}$ is a right eigenmatrix of the transfer operator with eigenvalue 1, and so spans the null space of the left-hand operator.  Hence, subtracting $\mathbb{I}$ times
\begin{align}
e_0&=\mathrm{Tr}\left[\rho_{\mathrm{half-infinite}}T_{L\hat{\sigma}_z}\left(A,T_{L\hat{\sigma}_z}\left(A,\mathbb{I}\right)\right)-hT_{L\hat{\sigma}_x}\left(A,\mathbb{I}\right)\right]\, ,
\end{align}
from the right hand side ensures that the right hand side is orthogonal to the null space of the left hand side and so this equation has a consistent solution in spite of being singular.  The extracted quantity $e_0$ is the energy density.  The singularity in the relation for $\mathbb{E}_1$ amounts to an ambiguity in defining the zero of energy.

The above observations lead directly to an algorithm for the energy density of a left-canonical iMPS in the thermodynamic limit, which for the TFIM takes the form of Algorithm \ref{alg:TFIMED}.
\begin{algorithm}[H]
\caption{Algorithm for the energy density of the TFIM with respect to a left-canonical iMPS \label{alg:TFIMED}}
\begin{algorithmic}
\Require $A$ in left canonical form, $\sum_i \mathbb{A}^{i\dagger} \mathbb{A}^i=\mathbb{I}$
\Function{EnergyDensity}{$A,h$}
\State $T_{\left(\alpha\alpha'\right),\left(\beta\beta'\right)} \gets \sum_{i}A_{\alpha \beta}^{i\star}A_{\alpha'\beta'}^{i}$
\State $\rho_{\mathrm{half-infinite}}\gets$ right eigenvector of $T$ with eigenvalue 1
\State $\mathbb{E}_z\gets \sum_{ii'}\langle i|\hat{\sigma}_z|i'\rangle \mathbb{A}^{i\dagger} \mathbb{A}^{i'}$
\State $\mathbb{E}_z\gets \sum_{ii'}\langle i|\hat{\sigma}_z|i'\rangle  \mathbb{A}^{i\dagger} \mathbb{E}_z \mathbb{A}^{i'}$
\State $\mathbb{E}_x\gets \sum_{ii'}\langle i|\hat{\sigma}_x|i'\rangle \mathbb{A}^{i\dagger} \mathbb{A}^{i'}$
\State $e_0\gets \mathrm{Tr}\left[\rho_{\mathrm{half-infinite}}\left(\mathbb{E}_z-h\mathbb{E}_x\right)\right]$
\State \Return $e_0$
\EndFunction
\end{algorithmic}
\end{algorithm}
This algorithm can hence be used as a cost function in a minimization routine to optimize the tensor $A$.  In order to ensure that the tensor remains in left-canonical form during the optimization, we utilize Riemannian manifold gradient descent with the matrix $\tilde{A}_{\left(\alpha i\right)\beta}=A_{\alpha\beta}^{i}$ forming an element of the Stiefel manifold $\mathrm{St}\left(\chi d,\chi\right)$~\cite{edelman1998geometry}.  As detailed in other recent works~\cite{hauru2021riemannian,luchnikov2020riemannian,wall2021tree}, this procedure consists of projection of cost function gradients onto the tangent space of the manifold at the current estimate point and subsequent update of the tensor using a retraction of this projected gradient.  Vector transport can be used to incorporate information from previous iterations through the momentum technique.  This process integrates readily with modern numerical packages performing automatic differentiation to obtain gradients, such as TensorFlow~\cite{abadi2016tensorflow} and PyTorch~\cite{NEURIPS2019_9015}.

The tensor output from this optimization procedure produces the proper half-infinite density matrix as the fixed point of the transfer matrix.  However, when practically implemented on quantum hardware we will produce an approximation to the half-infinite density matrix by using the ``burn in" procedure with a fixed number $N_b'$ of iterations (taken as a hyperparameter in our approach).  Hence, we will optimize the initial state of the bond qubits (boundary vector) to reproduce the true half-infinite density matrix obtained as the fixed point of the transfer operator as closely as possible using this procedure.  The cost function for this process is given as Algorithm \ref{alg:CFDM}.
\begin{algorithm}[H]
\caption{Cost function for difference between true half infinite density matrix and that obtained from a finite burn-in procedure \label{alg:CFDM}}
\begin{algorithmic}
\Require $A$ in left canonical form, $\sum_i \mathbb{A}^{i\dagger} \mathbb{A}^i=\mathbb{I}$
\Require $V$ with unit norm, $\sum_{\alpha} \left|V_{\alpha}\right|^2=1$
\Function{BurnInCost}{$A,V,N_b'$}
\State $T_{\left(\alpha\alpha'\right),\left(\beta\beta'\right)} \gets \sum_{i}A_{\alpha \beta}^{i\star}A_{\alpha'\beta'}^{i}$
\State $\rho_{\mathrm{half-infinite}}\gets$ right eigenvector of $T$ with eigenvalue 1
\State $B_{\alpha}^i \gets \sum_{\beta} A_{\alpha \beta}^iV_{\beta}$
\State $U_{\alpha \alpha '}\gets \sum_i B_{\alpha}^{i\star} B_{\alpha'}^i$
\State $N\gets 1$
\While{$N < N_b'$}
\State $Q_{\alpha\beta}^{i} \gets \sum_{\alpha'} U_{\alpha\alpha'}A_{\alpha' \beta}^i$
\State $U_{\alpha \alpha '}\gets \sum_{\beta,i} Q_{\beta \alpha'}A_{\beta \alpha}^{i\star}$
\State $N\gets N+1$
\EndWhile
\State \Return $\mathrm{norm}\left(\mathbb{U}-\rho_{\mathrm{half-infinite}}\right)$
\EndFunction
\end{algorithmic}
\end{algorithm}
This cost function can again be integrated with manifold optimization on the $\chi$-dimensional sphere manifold to optimize the boundary state $V$.  With the optimized tensor in hand, we can utilize the gauging and compilation procedures described in the main text to convert this into a quantum circuit for implementation on hardware.

\section{Algorithm for sampling a classical product state from a matrix product state}
\label{app:CPSSampling}

In this appendix, we review an efficient algorithm for sampling a classical product state in the basis defined by a collection of $L$ Hermitian operators $\{\hat{O}_i,i=0,\dots,L-1\}$ from a finite MPS $|\psi\left[A\right]\rangle$ on $L$ sites with open boundary conditions ($\langle l|=|r\rangle=\delta_{\alpha,0}$) according to its probability of occurrence in the MPS.  Specifically, product states $|\lambda_0\dots\lambda_{L-1}\rangle$ in a basis to be described will be sampled from $|\langle \psi\left[A\right]|\lambda_0\dots \lambda_{L-1}\rangle|^2$.  This algorithm first appeared in the context of minimally entangled typical thermal state (METTS) algorithms~\cite{stoudenmire2010minimally}.  Given the Hermitian operators $\hat{O}_i$ with spectral decomposition $\hat{O}_i=\sum_{\lambda_i}\lambda_i|\lambda_i\rangle\langle \lambda_i|$ we can define projection operators $\hat{P}_{i,\mu}=|\lambda_{i,\mu}\rangle\langle \lambda_{i,\mu}|$.  Now, assuming the MPS is in left-canonical form, we sample the $(L-1)^{\mathrm{th}}$ site by sampling from the categorical distribution defined by the vector $v_{\mu}=\mathrm{Tr}\left[\hat{P}_{\mu}\hat{\rho}_{L-1}\right]$, in which $\left[\hat{\rho}_{L-1}\right]_{j,j'}=\sum_{\alpha} A_{\alpha,0}^{[{L-1}] j}A_{\alpha,0}^{[{L-1}] j'\star}$ is the $(L-1)^{\mathrm{th}}$-site's reduced density matrix.  Given the index chosen by the random sampling, $\mu_{L-1}$, we now construct an MPS representation of $\hat{P}_{{L-1},\mu_{L-1}}|\psi\left[A\right]\rangle$ by replacing $A_{\alpha}^{[{L-1}]}\leftarrow \sum_{j} \langle j|\lambda_{\mu_{L-1}}\rangle A_{\alpha,0}^{[{L-1}],j}$.  Since the $(L-1)^{\mathrm{th}}$ site now no longer has a site index, we can absorb this tensor into the tensor for site $(L-2)$ as
\begin{align}
A_{\alpha,0}^{[L-2]j}&\leftarrow \sum_{\beta} A_{\alpha\beta}^{[L-2]j}A_{\beta}^{[{L-1}]}\, ,
\end{align}
to define an MPS on $(L-1)$ sites describing $\hat{P}_{{L-1},\mu_{L-1}}|\psi\left[A\right]\rangle$.  This process is now repeated, constructing the single-site reduced density matrix for site $(L-2)$, sampling from the (normalized) vector obtained by tracing the reduced density matrix with the projectors onto the eigenspaces, applying the chosen projector to the state, and contracting the MPS tensor from the sampled site into the neighboring MPS tensor.  Repeating for all remaining sites, the result of this algorithm is the string $(\mu_0,\dots,\mu_{L-1})$ which can be re-interpreted as a CPS $|\lambda_{\mu_0}\dots \lambda_{\mu_{L-1}}\rangle$.  This algorithm produces CPS sampled according to $\left|\langle \lambda_{\mu_0}\dots \lambda_{\mu_{L-1}}|\psi\left[A\right]\rangle\right|^2$, and has a $\mathcal{O}\left(\chi^2 L\right)$ cost that is subleading to the $\mathcal{O}\left(L\chi^3\right)$ cost of variational ground state search.

\section{Online algorithm for determining the number of shots needed for accurate classification}
\label{app:Bayes}
For classically preconditioned models such as those described in this paper, the Wilson score interval can be used to estimate the number of experimental shots required to infer the class of a given input state $|\mathbf{x}\rangle$ with a desired accuracy, as $p_0$ can be obtained from the classically tractable model.  As models are scaled up towards the classically intractable regime, we no longer have knowledge of $p_0$, and it must instead be inferred from the experimental data.  In this scenario, it is useful to have an \emph{online} algorithm which estimates a distribution over $p_0$ conditioned on the shot history, such that statistical measures on this conditional distribution can determine when a sufficient number of shots have been taken.  A natural setting for this is Bayesian inference.  Using a uniform prior distribution over $p_0$, corresponding to the absence of any a priori knowledge about $p_0$, and the assumption that each shot of the experiment is a Bernoulli trial and hence described by a binomial distribution, the resulting posterior distribution for $p_0$ conditioned on a shot history of $n_0$ zeros and $n_1$ ones is a beta distribution
\begin{align}
P\left(p_0|n_0,n_1\right)&=\frac{1}{B\left(n_0+1,n_1+1\right)}p_0^{n_0}\left(1-p_0\right)^{n_1}\, ,
\end{align}
in which $B\left(x,y\right)$ is the beta function.  Hence, the probability that $p_0\le \frac{1}{2}$, corresponding to the posterior model predicting class one, is given by the cumulative distribution function of this posterior distribution evaluated at $\frac{1}{2}$.  For the present case, this is
\begin{align}
p_{<}\equiv \int_0^{1/2}\! dp_0\,P\left(p_0|n_0,n_1\right)&=I_{\frac{1}{2}}\left(n_0+1,n_1+1\right)\, ,
\end{align}
in which $I_x\left(a,b\right)$ is the regularized incomplete beta function.  With this, one can set a threshold $\frac{1}{2}<p_{\star}<1$ such that no more shots are taken and the class is called as one when $p_<>p_{\star}$ or zero when $p_<<1-p_{\star}$.  These criteria define that the interval $[0,1/2]$ is a credible interval with probability $p_{\star}$, in the Bayesian sense.

Using the identities
\begin{align}
I_x\left(a+1,b\right)&=I_x\left(a,b\right)-\frac{x^a\left(1-x\right)^b}{aB\left(a,b\right)}\, ,\\
I_x\left(a,b+1\right)&=I_x\left(a,b\right)-\frac{x^a\left(1-x\right)^b}{bB\left(a,b\right)}\, ,\\
B\left(a+1,b\right)&=B\left(a,b\right)\frac{a}{a+b}\, ,\\
B\left(a,b+1\right)&=B\left(a,b\right)\frac{b}{a+b}\, ,
\end{align}
and $I_{\frac{1}{2}}\left(1,1\right)=\frac{1}{2}$, $B\left(1,1\right)=1$, we can develop an online Bayesian algorithm that will halt when the requisite number of shots have been taken, summarized in Algorithm \ref{alg:Bayes}.
\begin{algorithm}[H]
\caption{\label{alg:Bayes} Online algorithm to determine class from probabilistic classifier with confidence parameter $p_{\star}$}
\begin{algorithmic}
\Require $p_{\star}$, method to obtain probabilistic class prediction \texttt{SamplePrediction}
\Function{ClassFromSamples}{$p_{\star}$}
\State $B\gets 4$, $n_0\gets 0$, $n_1\gets 0$, $p_<\gets \frac{1}{2}$
\While{True}
\State $s\gets $\texttt{SamplePrediction}
\If{$s$ == 0}
\State $p_<\gets p_<-\frac{1}{\left(n_0+1\right)B}$
\State $B\gets 2B\frac{n_0+1}{n_0+n_1+2}$
\State $n_0\gets n_0+1$
\Else
\State $p_<\gets p_<+\frac{1}{\left(n_1+1\right)B}$
\State $B\gets 2B\frac{n_1+1}{n_0+n_1+2}$
\State $n_1\gets n_1+1$
\EndIf
\If{$p_<>p_{\star}$}
\Return 1
\ElsIf{$p_<<1-p_{\star}$}
\Return 0
\EndIf
\EndWhile
\EndFunction
\end{algorithmic}
\end{algorithm}
From this algorithm, we can also obtain the maximum a posteriori estimate of $p_0$ as $\frac{n_0}{n_0+n_1}$.

\bibliography{Refs}
\bibliographystyle{apsrev4-1}

\end{document}